\documentclass[aps,showpacs,prl,reprint,superscriptaddress,longbibliography]{revtex4-1}

\usepackage[colorlinks=true,linkcolor=blue,citecolor=blue,urlcolor=blue]{hyperref}

\usepackage{setspace} 
\usepackage{graphicx}
\usepackage{amsmath}
\usepackage{color}
\usepackage{amsmath}
\usepackage{amssymb}
\usepackage{verbatim}
\usepackage{latexsym}
\usepackage{enumerate} 
\usepackage{bm} 
\usepackage[caption=false]{subfig}
\begin{document}

\title{Structure and metallicity of phase V of hydrogen}

\author{Bartomeu Monserrat} 
\email{bm418@cam.ac.uk}
\affiliation{Department of Physics and Astronomy, Rutgers University,
  Piscataway, New Jersey 08854-8019, USA} 
\affiliation{TCM Group,
  Cavendish Laboratory, University of Cambridge, J.\ J.\ Thomson
  Avenue, Cambridge CB3 0HE, United Kingdom}

\author{Neil D. Drummond}
\affiliation{Department of Physics, Lancaster University, Lancaster
  LA1 4YB, United Kingdom}

\author{Philip Dalladay-Simpson} 
\affiliation{Center for High Pressure Science and Technology Advanced Research, Shanghai, People's Republic of China}

\author{Ross T.\ Howie} 
\affiliation{Center for High Pressure Science and Technology Advanced Research, Shanghai, People's Republic of China}

\author{Pablo L\'opez R\'ios}
\affiliation{TCM Group,
  Cavendish Laboratory, University of Cambridge, J.\ J.\ Thomson
  Avenue, Cambridge CB3 0HE, United Kingdom}
  \affiliation{Max-Planck Institute for Solid State Research,
               Heisenbergstra{\ss}e 1, 70569 Stuttgart, Germany}

\author{Eugene Gregoryanz} 
\affiliation{Center for High Pressure Science and Technology Advanced Research, Shanghai, People's Republic of China}
\affiliation{Centre for Science at Extreme Conditions and School of Physics and Astronomy, University of Edinburgh, Edinburgh EH9 3JZ, United Kingdom} 
\affiliation{Key Laboratory of Materials Physics, Institute of Solid State Physics, Chinese Academy of Sciences, Hefei, People's Republic of China}

\author{Chris J.\ Pickard} 
\affiliation{Department of Materials
  Science and Metallurgy, University of Cambridge, 27 Charles Babbage
  Road, Cambridge CB3 0FS, United Kingdom} 
\affiliation{Advanced
  Institute for Materials Research, Tohoku University 2-1-1 Katahira,
  Aoba, Sendai, 980-8577, Japan}

\author{Richard J.\ Needs} 
\affiliation{TCM Group, Cavendish
  Laboratory, University of Cambridge, J.\ J.\ Thomson Avenue,
  Cambridge CB3 0HE, United Kingdom}

\date{\today}


\begin{abstract}
  A new phase V of hydrogen was recently claimed in experiments above
  $325$~GPa and $300$~K. Due to the extremely small sample size at
  such record pressures the measurements were limited to Raman
  spectroscopy.  The experimental data on increase of pressure shows
  decreasing Raman activity and darkening of the sample, which
  suggests band-gap closure and impending molecular dissociation, but
  no definite conclusions could be reached.  Furthermore, the
  available data is insufficient to determine the structure of phase
  V, which remains unknown.  Introducing saddle-point \textit{ab
    initio} random structure searching (sp-AIRSS), we find
  several new structural candidates of hydrogen which could describe
  the observed properties of phase V. We investigate hydrogen
  metallisation in the proposed candidate structures, and demonstrate
  that smaller band gaps are associated with longer bond lengths. We
  conclude that phase V is a stepping stone towards metallisation.
\end{abstract}

\maketitle

The study of dense hydrogen is important to fundamental physics and 
astrophysics~\cite{hydrogen_superconductivity,silvera_h_review_1980,ceperley_rmp_h_he,h_exp_review}.
Currently the most interesting question relates to the
metallisation and dissociation of molecular hydrogen under pressure,
which has not yet been achieved in the solid state, even though it was
first proposed in
$1935$~\cite{wigner_huntington_metallic_hydrogen_1935}.  The known
phases I, II, III, and IV/IV$^{\prime}$ of solid hydrogen, which have
been characterised extensively 
experimentally~\cite{h_phase_II,h_phase_III,h_phase_IV_eremets,h_phase_IV_gregoryanz,h_phase_IVp_gregoryanz}
and
theoretically~\cite{nature_physics_h,ceperley_hydrogen_superconductivity_2011,phase_iv_prb,phase_iv_prb_erratum,ma_cc_hydrogen,md_raman_ackland,prl_dissociation_hydrogen,h_dissociation_morales,hydrogen_nature_communications},
exhibit molecular bonds and are insulating.

Dalladay-Simpson and co-workers recently reported Raman spectroscopy experiments on H$_2$, D$_2$, and HD up to pressures of $388$~GPa at $300$~K~\cite{gregoryanz_nature_phase_V}. In these experiments, they identified a new phase V of H$_2$ and HD above $325$~GPa and at $300$~K, which was suggested to be at the onset of dissociation and could therefore represent a stepping stone towards full metallisation. Several experimental reports followed, claiming metallisation of H$_2$ under different pressure-temperature conditions~\cite{silvera_h2pre_science,eremets_semimetallic_arxiv}, but the validity of these experiments is yet to be confirmed~\cite{gregoryanz_criticism_of_silvera,goncharov_criticism_of_silvera}. In this Letter, we focus on phases IV, IV$^{\prime}$, and V as described in Refs.~\cite{h_phase_IVp_gregoryanz,gregoryanz_nature_phase_V}.

On the theoretical front, a number of candidate structures have been
proposed to explain the observed experimental phases of high-pressure
hydrogen up to
$300$~GPa~\cite{nature_physics_h,phase_iv_prb,phase_iv_prb_erratum,hexagonal_phase_III_monserrat}. Of
these, the monoclinic $C2/c$ structure is currently the best candidate
for phase III around
$300$~GPa~\cite{nature_physics_h,hexagonal_phase_III_monserrat}, as it
exhibits Raman and infra-red (IR) spectra consistent with those observed
experimentally. The monoclinic $Pc$ structure is the best candidate
for phase IV~\cite{phase_iv_prb,phase_iv_prb_erratum} due to its mixed layered nature that
leads to the two vibron peaks observed experimentally. Recent quantum
Monte Carlo and free energy calculations have confirmed these phases
to be energetically favourable in the pressure range in which phases
III and IV are observed~\cite{hydrogen_nature_communications}. The
most stable atomic hydrogen candidate structure is tetragonal and has space group
$I4_1/amd$~\cite{nature_physics_h,ceperley_hydrogen_superconductivity_2011,prl_dissociation_hydrogen,h_dissociation_morales}.
Despite the large number of candidate structures known for
high-pressure hydrogen, none provides a good model for the recent
experimental observations at pressures above about $300$~GPa. 

Discovering candidate structures using searching methods has been
successful in many systems, particularly at high
pressure~\cite{PhysRevLett.97.045504,genetic_oganov,particle_swarm_ma,Pickard2011,zurek_method}.
As an example, the lowest-enthalpy candidate structures for phases II,
III, and IV of hydrogen have been found using the \textit{ab initio} random
structure searching (AIRSS) method~\cite{nature_physics_h,phase_iv_prb,phase_iv_prb_erratum}. The experimental discovery
of phase V, for which there is no obvious candidate structure, prompts
the question of whether it is necessary to go beyond current structure
searching methods in this case.

Standard structure searching methods such as AIRSS are restricted to
structures associated with minima of the potential energy
landscape. However, thermodynamically stable structures associated
with saddle points that are dynamically stabilised by anharmonic
nuclear motion are known to
exist~\cite{lines_glass_perovskites_book,RevModPhys.84.945}. The
high-temperature cubic perovskite phase of BaTiO$_3$ provides a
well-known
example~\cite{lines_glass_perovskites_book,batio3_vanderbilt}.

A variety of computational methods has been used to determine the
dynamical stability of such structures, including Monte
Carlo~\cite{batio3_vanderbilt}, molecular
dynamics~\cite{md,ab_initio_md}, path integral molecular
dynamics~\cite{pimd1,pimd2}, and local anharmonic vibrational
methods~\cite{PhysRevLett.100.095901,PhysRevB.84.180301,PhysRevB.86.054119,PhysRevB.87.144302,errea_prb}.
These methods can determine the dynamical stability of a known
saddle-point structure but they have not been
used to find previously unknown saddle-point structures.  The
following question arises: can we devise a systematic approach to
searching for previously unknown structures associated with saddle
points of the energy landscape? The large nuclear effects of hydrogen
make it an ideal system in which to explore this possibility.

We address these questions using saddle-point \textit{ab initio} random
structure searching (sp-AIRSS). 
Saddle-point structures stabilised by anharmonic nuclear
motion are typically of higher symmetry than their
\textit{broken-symmetry} counterparts.  Based on this observation, we
use sp-AIRSS to impose high-symmetry constraints during structure
searches.  For example, imposing cubic symmetry on BaTiO$_3$, leads to
the known cubic phase, but removing the symmetry constraints leads
instead to the rhombohedral phase. The symmetry constraints bias the
search towards the high-symmetry structures that are expected to be
stable when the vibrational amplitudes are large. 
We emphasise that this strategy enables the discovery of structures which cannot be found in unconstrained searches because correspond to minima of the free energy landscape but not of the static lattice energy landscape.
 We then remove the
symmetry constraints and relax the reference structure using an
anharmonic vibrational method. In this work we have used the
vibrational self-consistent field method of Ref.~\cite{PhysRevB.87.144302}, but
any of the available anharmonic methods may be applicable at this stage of the
calculation~\cite{batio3_vanderbilt,md,ab_initio_md,pimd1,pimd2,PhysRevLett.100.095901,PhysRevB.84.180301,PhysRevB.86.054119,PhysRevB.87.144302,errea_prb}. The
structure may then relax to a minimum or saddle point of the potential
energy landscape.

The lowest-enthalpy known hydrogen structures have monoclinic symmetry
with space groups $C2/c$ (model for phase III) and $Pc$ (model for
phase IV)~\cite{nature_physics_h,phase_iv_prb,phase_iv_prb_erratum}. To search for new
candidate structures we have therefore performed sp-AIRSS searches
imposing space groups of orthorhombic or higher symmetry. The searches
have led to the discovery of three new energetically competitive
structures at pressures for which phase V is observed. These
structures have orthorhombic symmetry with space groups $Pca2_1$,
$Pna2_1$, and $Pcaa$, and $48$ atoms in the primitive cell. $Pca2_1$
and $Pna2_1$ are mixed layered structures similar to $Pc$ in which
alternate layers exhibit shorter and longer molecular bond lengths,
resulting in two vibron peaks in the Raman and IR spectra. $Pcaa$ has
a single type of layer.

Our analysis in this work is based on these three new structures,
together with the previously reported structures
$C2/c$~\cite{nature_physics_h}, $Cmca$-$4$ and
$Cmca$-$12$~\cite{nature_physics_h} (where $4$ and $12$ indicate the
number of atoms in the primitive cell), $Pc$~\cite{phase_iv_prb,phase_iv_prb_erratum} and
$Ibam$~\cite{nature_physics_h}. The $C2/c$ and $Cmca$ structures model
phase III and 
all theoretical methods predict that $C2/c$ is more
stable at lower pressures and $Cmca$ at higher pressures, but the
precise pressure above which $Cmca$ becomes stable is highly-dependent
on the level of theory used. An hexagonal structure of space group
$P6_122$ has recently been proposed as a candidate for phase III at
pressures below $200$~GPa~\cite{hexagonal_phase_III_monserrat}, but in
this work we focus on higher pressures, and therefore do not include
it in our analysis. The $Ibam$ structure is an extreme member of the
family of mixed structures, in which the weakly-bound layer is
graphene-like and molecular bonds are no longer present.

Of all structures considered, $Pna2_1$ and $Pca2_1$ are dynamically
unstable at the harmonic vibrational level, and their broken-symmetry
counterpart is a monoclinic structure. $Ibam$ is also dynamically
unstable, while the rest are dynamically stable. Note that unless
sp-AIRSS had been used, $Pca2_1$ and $Pna2_1$ would have fallen into
the corresponding broken-symmetry monoclinic structure, and would have
gone unnoticed. The symmetry constraints prevent this and allow the
potential discovery of new structures stabilised by anharmonic
vibrations.

\begin{figure*} 
\centering
\subfloat[][(a) BLYP functional.]{
\includegraphics[scale=0.32]{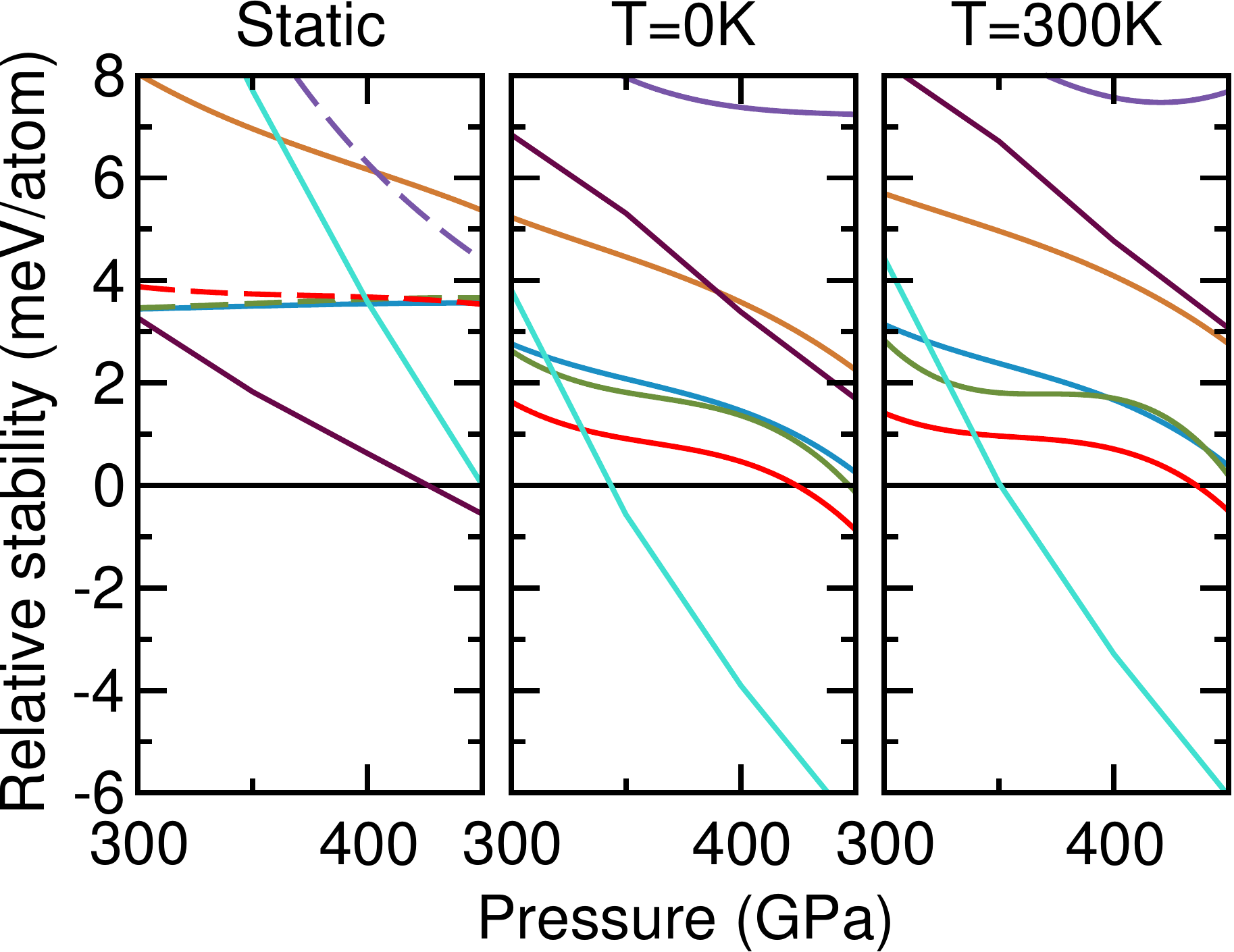}
\label{subfig:blyp}
}
\hspace{0.1cm}
\subfloat[][(b) PBE functional.]{
\includegraphics[scale=0.32]{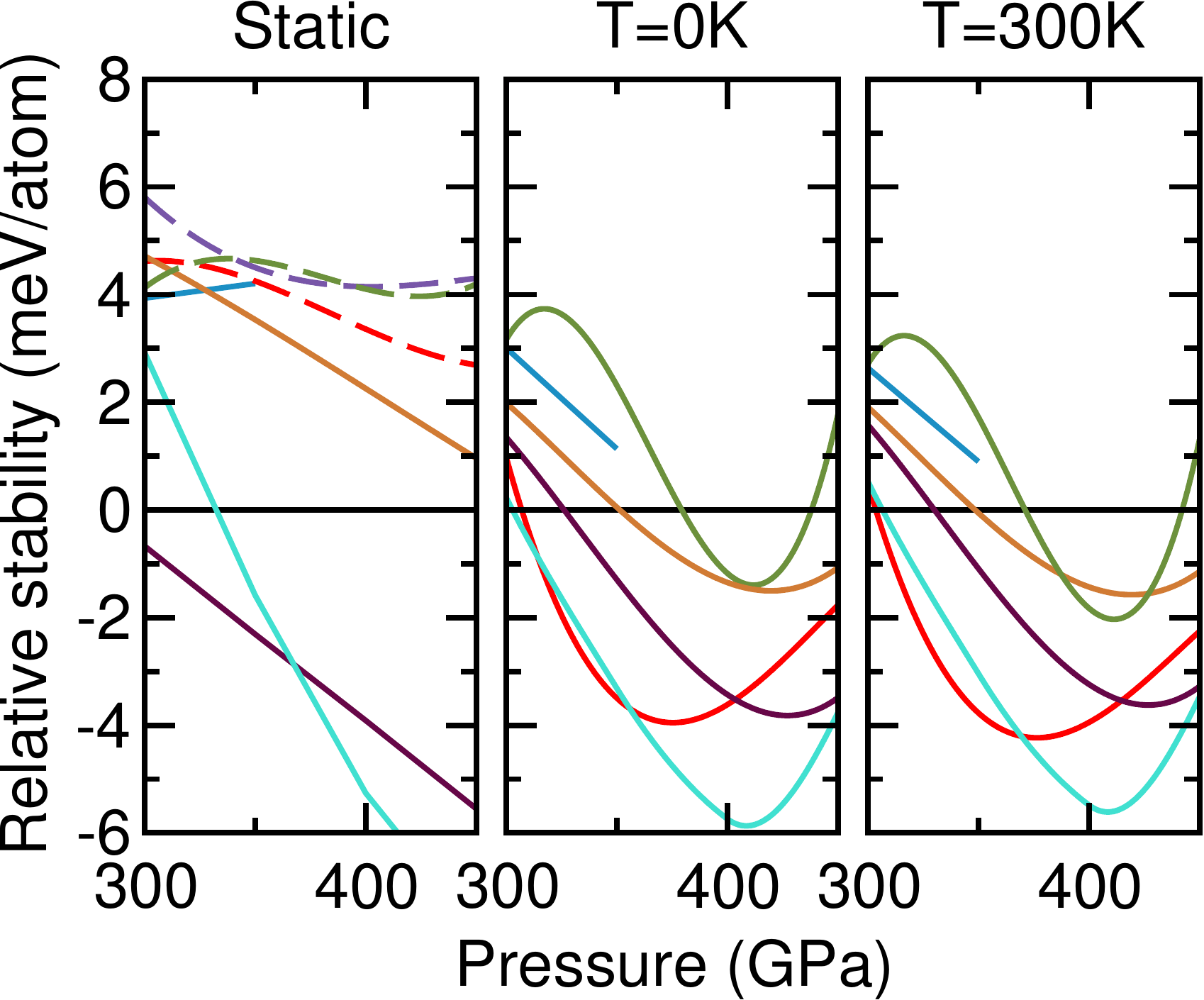}
\label{subfig:pbe}
}
\hspace{0.1cm}
\subfloat[][\hspace{-1.5cm}(c) DMC.]{
\includegraphics[scale=0.32]{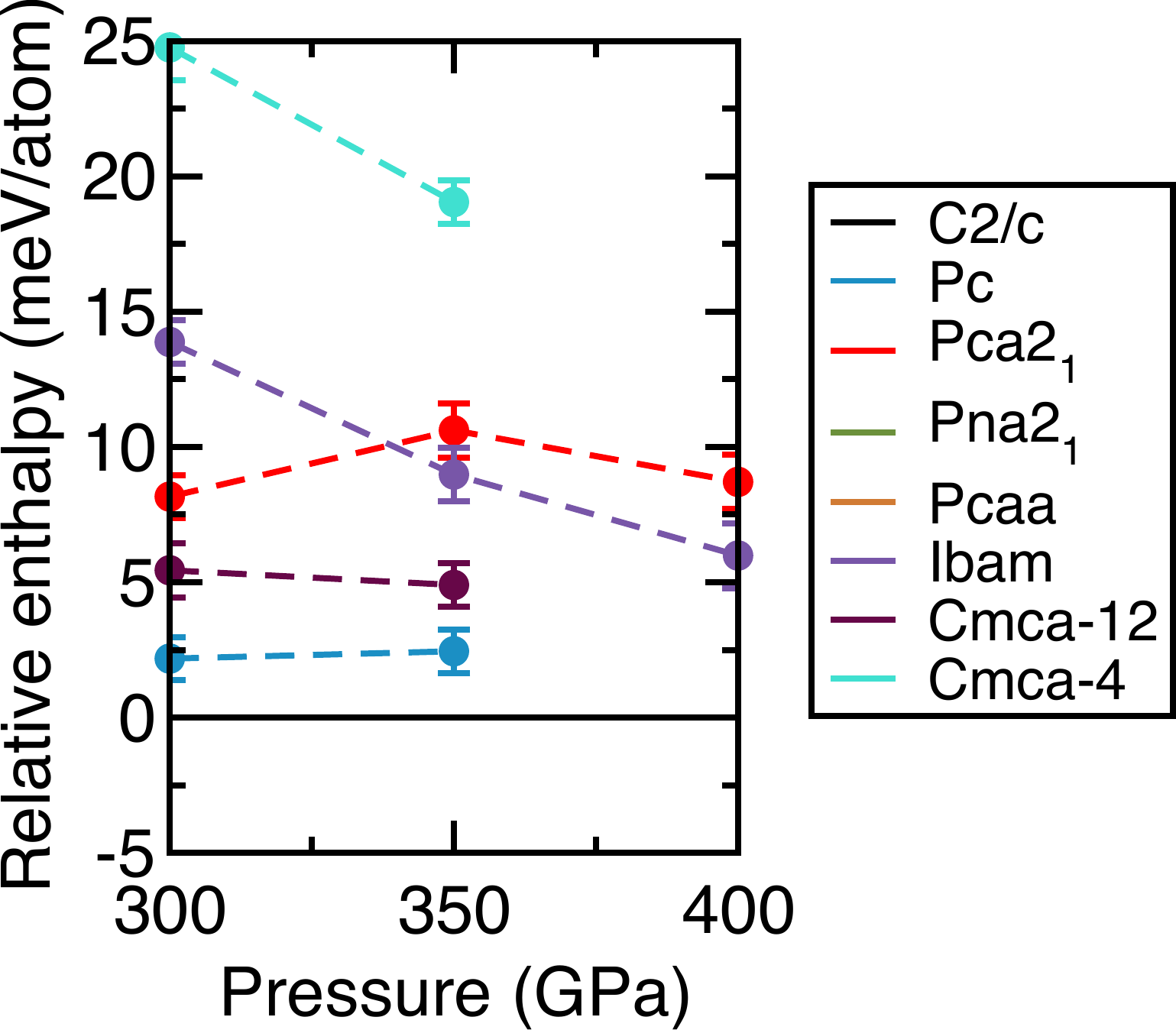}
\label{subfig:qmc}
}
\caption{Relative enthalpies using the (a) BLYP and (b) PBE DFT functionals, and using (c) DMC. The DFT results are at the static lattice level, at $T=0$~K (including zero-point motion), and at $T=300$~K, and the dashed lines in the static lattice diagrams indicate enthalpies corresponding to structures at saddle-points of the energy landscape. The DMC results are at the static lattice level, and the dashed lines between the DMC points are a guide to the eye only.}
\label{fig:enthalpy}
\end{figure*}

We have used first-principles methods based on density functional
theory (DFT) as implemented in the {\sc castep}~\cite{CASTEP} code to
calculate the relative stability of the eight structures under
consideration. We have used both the BLYP exchange-correlation
functional~\cite{blyp_exchange,blyp_correlation}, which has been shown
to be accurate for the description of molecular hydrogen
structures~\cite{clay_benchmarking}, and the PBE exchange-correlation
functional~\cite{PhysRevLett.77.3865}, which we find to favour atomic
phases compared to the BLYP functional. Due to the small energy
differences between competing structures of only a few meV, the
resulting phase diagrams are sensitive to the level of theory
used~\cite{azadi_h_dft_qmc,clay_benchmarking,hydrogen_nature_communications}.
We therefore also perform selected diffusion Monte Carlo (DMC) calculations
using the {\sc casino} package~\cite{qmc_jphys_review} to establish the
validity of our conclusions based on the DFT results.
To calculate the
vibrational contribution to the energy including anharmonic
contributions we use the method of
Refs.~\cite{PhysRevB.87.144302,non_diagonal}. Further details of the
first principles calculations are provided in the Supplemental
Material~\cite{supplementary_spairss}.

In Figs.~\ref{subfig:blyp} and \ref{subfig:pbe} we report static lattice enthalpies,
zero-temperature enthalpies (including quantum zero-point motion), and
Gibbs free energies at $300$~K relative to $C2/c$ using DFT. The static lattice
enthalpies of $Pca2_1$, $Pna2_1$, and $Ibam$ are shown as dashed lines
to indicate dynamical instability at the harmonic level corresponding
to saddle points of the potential energy landscape. All three
structures become dynamically stable when lattice vibrations are
included. We also show selected static lattice DMC calculations 
in Fig.~\ref{subfig:qmc}.

For both BLYP and PBE calculations, we observe that the $Cmca$-$4$
structure is the lowest in energy at the higher pressures
studied. This is consistent with earlier DFT studies, but we note that
using more accurate DMC calculations de-estabilises
this structure and removes it from the phase
diagram (see Fig.~\ref{subfig:qmc}). The $Cmca$-$12$ structure is also
de-estabilised within DMC, although to a smaller degree than the $Cmca$-$4$ 
structure.

The BLYP results show that, of the mixed layered structures, $Pca2_1$
is the most competitive energetically at both zero and $300$~K,
becoming more stable than $C2/c$ at pressures of about $420$~GPa. The
PBE results also favour $Pca2_1$ as the most stable mixed layered
structure, but it becomes more stable than $C2/c$ at significantly
lower pressures of about $300$~GPa, consistently with the observation
that PBE favours atomic phases compared to molecular phases ($Pca2_1$
has alternate layers with longer bond lengths than those observed in
$C2/c$). We also note that, at the PBE level, $Pc$ does not exist
above about $375$~GPa, as it falls into the $Cmca$-$4$
structure. Finally, we note that the $Pcaa$ structure, which is not
energetically competitive at the BLYP level, becomes more competitive
at the PBE level, a fact that we again attribute to the longer bond
lengths exhibited by $Pcaa$ when compared to $C2/c$.
Our static DMC calculations combined by the DFT vibrational energy estimates
confirm that $Pca2_1$ remains energetically competitive as a candidate structure of high pressure hydrogen (see Supplemental Material~\cite{supplementary_spairss}).

\begin{figure} \centering
\subfloat[][]{
\includegraphics[scale=0.32]{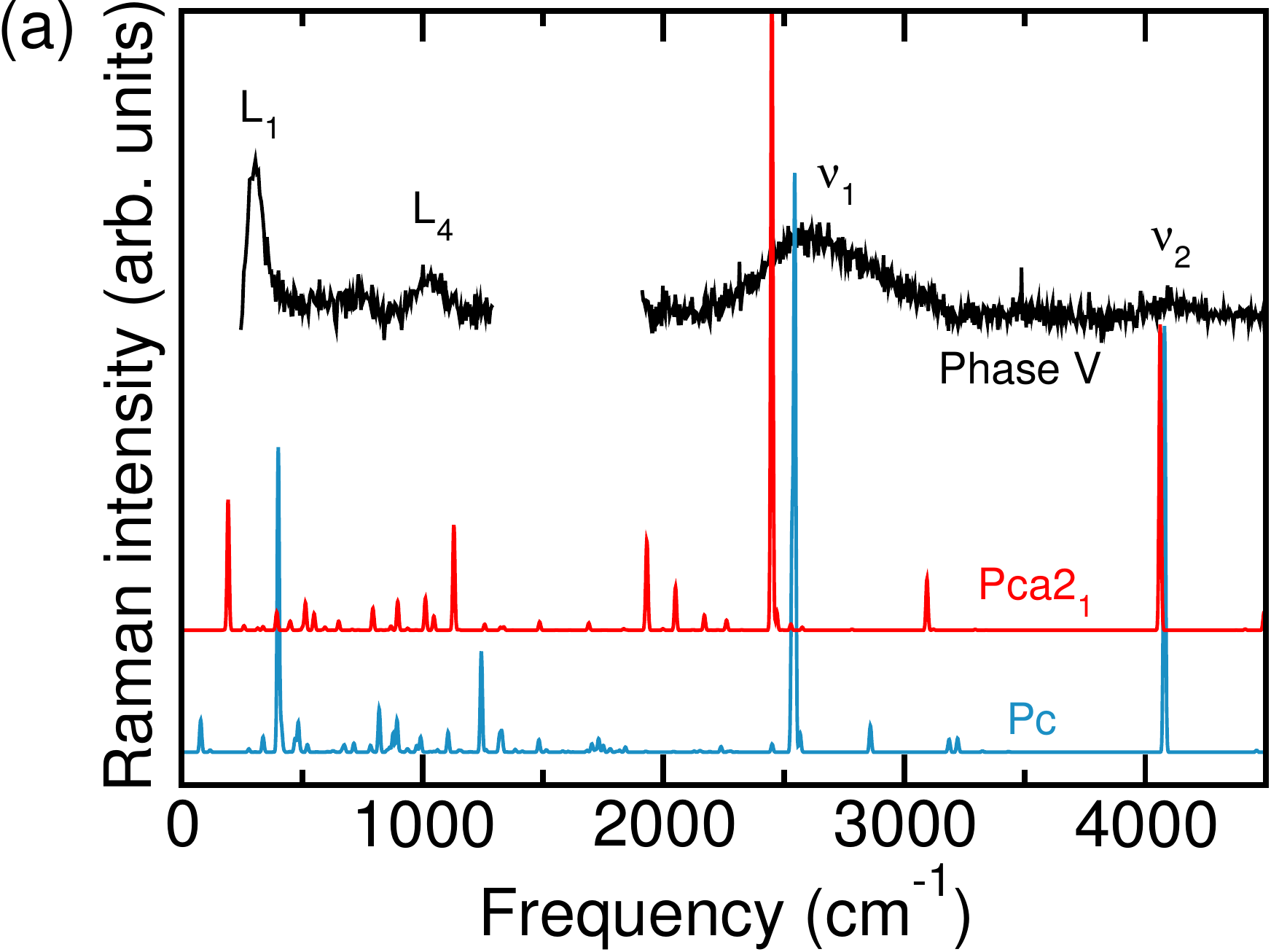}
\label{subfig:intensity}}  \\
\subfloat[][]{
\includegraphics[scale=0.32]{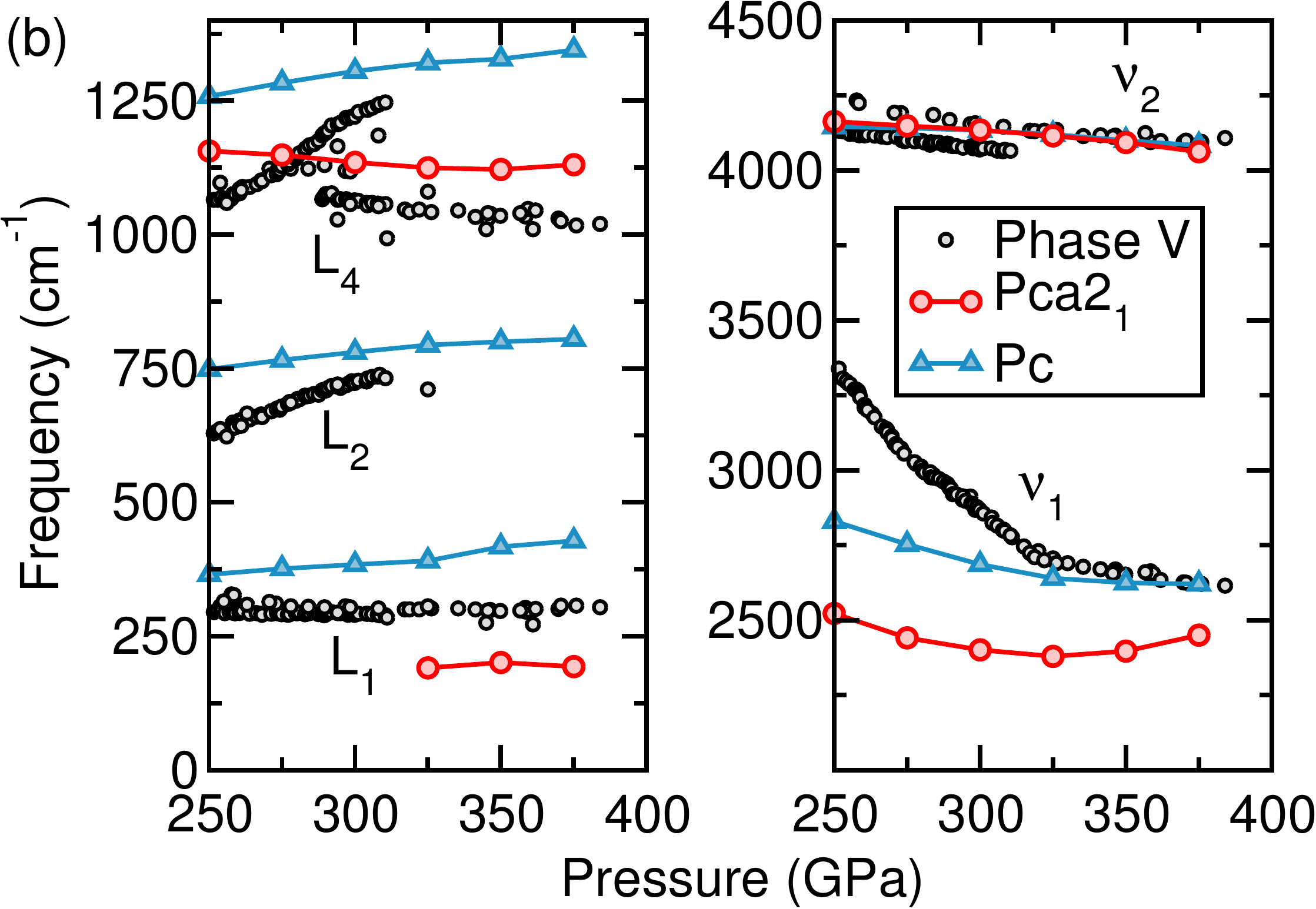}
\label{subfig:pressure}}  
\caption{(a) Raman spectra of $Pc$, $Pca2_1$, and phase V at
  $374$~GPa. The absence of data in the range $1500$--$1900$~cm$^{-1}$
  arises from the strong signal from the diamonds at these
  frequencies. (b) Pressure dependence of the frequencies of the most
  intense Raman peaks of $Pc$, $Pca2_1$, and phase V.}
\label{fig:raman}
\end{figure}

The experimental Raman spectrum of phase V is compared to the
theoretical harmonic spectra of $Pc$ and $Pca2_1$ calculated using the
PBE functional in Fig.~\ref{fig:raman}. Figure~\ref{subfig:intensity}
shows a comparison of the Raman intensities at $374$~GPa. In the
high-frequency regime, the frequency of the experimental $\nu_2$
vibron agrees with those of $Pc$ and $Pca2_1$. The frequency of the
$\nu_1$ vibron is marginally better reproduced by $Pc$ than by
$Pca2_1$. We also note that Magd\u{a}u and Ackland showed
that anharmonic contributions push the $\nu_2$ vibron to higher
energies in $Pc$~\cite{md_raman_ackland}, and a similar behaviour in
$Pca2_1$ would bring the latter into better agreement with
experiment. At the low-frequency regime the L$_1$ and L$_4$ modes of
phase V are in better agreement with $Pca2_1$ than with $Pc$. The
L$_2$ mode, which disappears upon entering phase V, is present in $Pc$
but missing in $Pca2_1$.

The pressure dependence of the Raman peaks is shown in
Fig.~\ref{subfig:pressure}, with phase IV below $325$~GPa, and phase V
at higher pressures. The pressure dependence of the $\nu_2$ vibron is
well-reproduced by both $Pc$ and $Pca2_1$. The frequency of the
low-energy vibron has a pressure dependence of $-1.4$~cm$^{-1}$/GPa in
phase V above $325$~GPa, which is much weaker than that of phase IV at
lower pressures (note the change in slope for $\nu_1$ around
$325$~GPa). The pressure dependence of $\nu_1$ in $Pc$ and $Pca2_1$ is
too weak at pressures below $325$~GPa, suggesting that they are not
good candidates for phase IV. However, we note that, as discussed
earlier, anharmonic effects significantly affect this
frequency~\cite{md_raman_ackland}, and therefore we cannot discard
these structures as candidates for phase IV.
The pressure dependence of the low frequency part of the Raman
spectrum of phase V is better-reproduced by $Pca2_1$ than by
$Pc$. 

A striking feature of the experimental Raman spectrum is the dramatic
increase in the width of the L$_1$ peak upon entering phase V, whose
FWHM increases from about $70$~cm$^{-1}$ at $325$~GPa to about
$160$~cm$^{-1}$ at $388$~GPa. The experimental data show that the
increase in the peak width is strongly isotope
dependent~\cite{gregoryanz_nature_phase_V}, suggesting a nuclear
origin for this feature. Therefore, it could be attributed to a
harmonic dynamical instability like the one exhibited by the $Pca2_1$
and $Pna2_1$ structures.

Our Raman spectra analysis suggests that the $Pca2_1$ structure is
consistent with phase V. The Raman spectrum of $Pna2_1$ is almost
identical to that of $Pc$, and both give poorer agreement with
experiment than $Pca2_1$. The $C2/c$, $Cmca$, and $Pcaa$ structures
cannot describe phase V, as they have a unique type of bond and thus a
single vibron. $Ibam$ is also an unlikely candidate for phase V, as
its vibron $\nu_1$ has a frequency below $2250$~cm$^{-1}$ in the
pressure range where phase V is observed. Details of the Raman spectra
of these phases are provided in the Supplemental
Material~\cite{supplementary_spairss}.

Overall, $Pca2_1$ is energetically competitive at the pressures at
which phase V has been observed, and crucially, of all structures considered,
only its Raman spectrum is consistent with that of phase V.
More generally, the known $Pc$ structure and the
new $Pna2_1$ and $Pca2_1$ structures are plausible candidates for the
high pressure hydrogen structures characterised by two strong vibrons,
that is, phases IV, IV$^{\prime}$, and V.

\begin{figure} \centering
\subfloat[][]{
\includegraphics[scale=0.38]{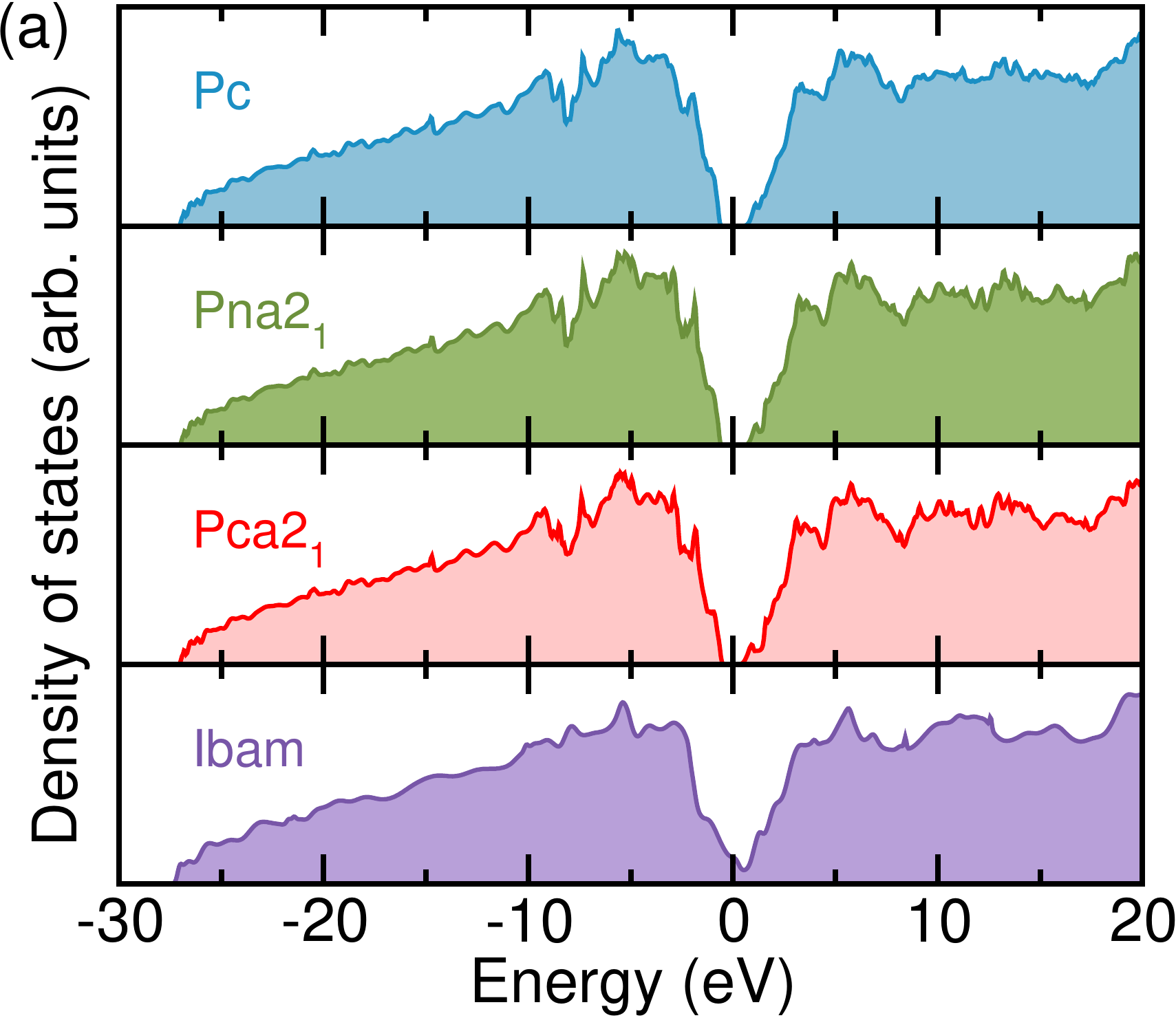}
\label{subfig:dos}}  \\
\vspace{-0.8cm}
\subfloat[][]{
\includegraphics[scale=0.38]{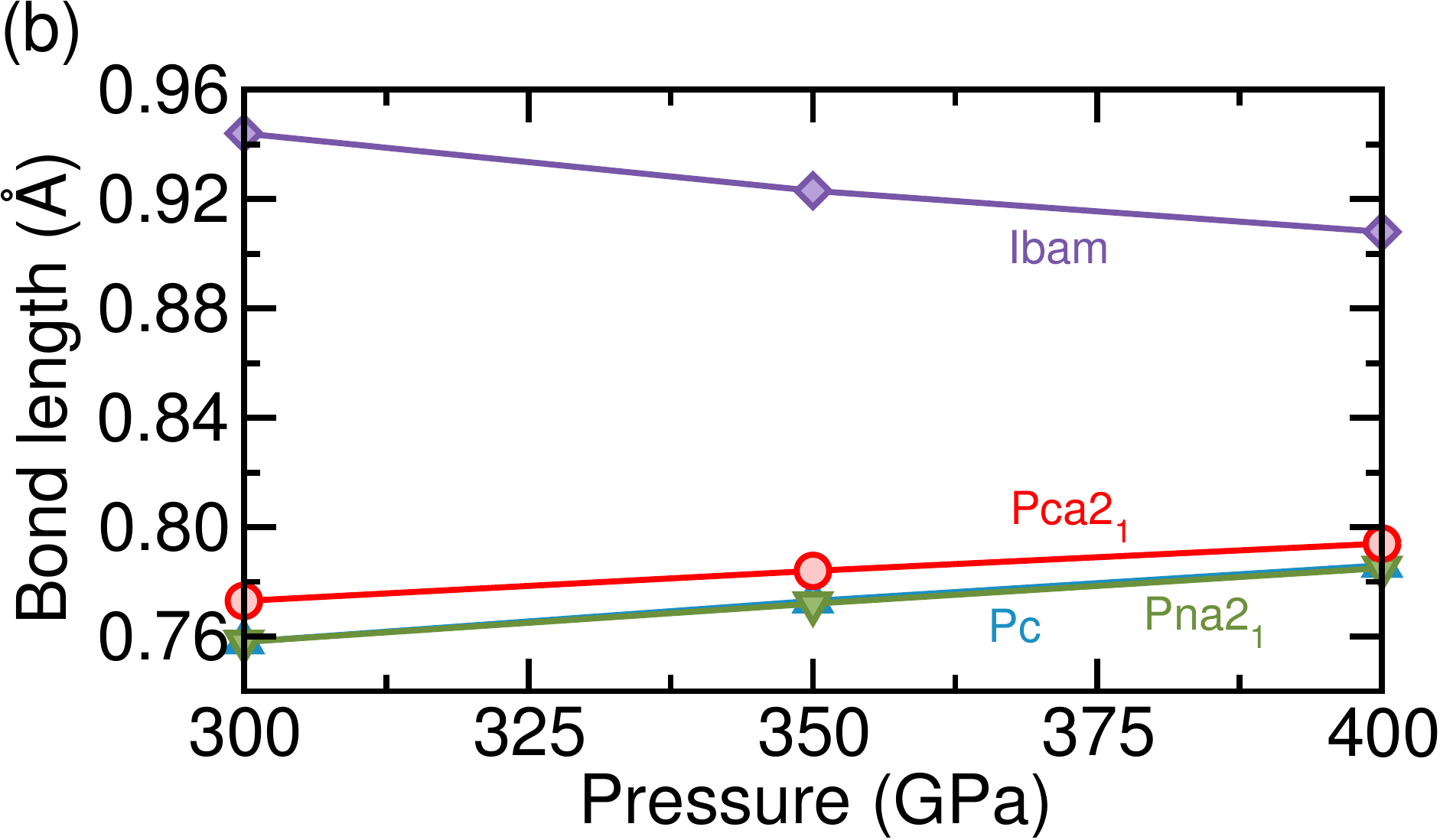}
\label{subfig:bond}} 
\caption{(a) Electronic densities of states of hydrogen candidate
  mixed layered structures $Pc$, $Pna2_1$, $Pca2_1$, and $Ibam$ at a
  pressure of $350$~GPa. (b) Static bond lengths of the $Pc$,
  $Pna2_1$, $Pca2_1$, and $Ibam$ structures for the layers with longer
  bonds. The bond lengths of $Pc$ and $Pna2_1$ are
  indistinguishable.} 
\label{fig:dos}
\end{figure}

Having discovered good candidate structures, we study the metallicity
of phase V. The study of band gap closure and metallisation in high
pressure hydrogen is a challenging problem. Band gaps are typically
underestimated by several electronvolts by Kohn-Sham 
DFT~\cite{azadi_hydrogen_gw_gap,azadi_h_metal_dmc_static}, whereas 
the neglect of electron-phonon coupling contributions tends to 
lead to an overestimation of the gap size~\cite{ceperley_h_elph_coupling,azadi_gap_pimd}.  
These two effects alter the gap in opposite directions, cancelling to some
extent. 
We consider static lattice DFT band structures, which contain valuable
insights on trends amongst the different structures, but cannot be used
reliably to estimate the actual band gap values. 

Metallisation in layered hydrogen structures has been proposed to
arise from the weakly bound layers that can be described as distorted
graphene sheets~\cite{cohen_graphene_hydrogen}. Here, we extend this
analysis to mixed-layered structures, where layers with short and long 
bond lengths coexist. In Fig.~\ref{subfig:dos} we show
the electronic densities of states at $350$~GPa for the four mixed
layered structures considered in this work. $Pc$, $Pna2_1$, and
$Pca2_1$ are all insulating. Of these four structures, $Pca2_1$ has
the smallest band gap by about $0.3$~eV.  This is a consequence of the
longer bond length in the weakly bound layers, as shown in
Fig.~\ref{subfig:bond}. The bond lengths of all of these structures in
the strongly-bound layers are comparable. 

Molecular dissociation is more pronounced in $Ibam$, as the weakly
bound layers are graphene-like, and the molecular character is
lost. This is shown by the longer bond lengths exhibited by $Ibam$ in
the graphene sheets (Fig.~\ref{subfig:bond}). The frequency of the
$\nu_1$ vibron in $Ibam$ increases with pressure, as expected from the
decreasing bond length of the graphene sheets.  In contrast, the
increase in bond length with pressure in $Pc$, $Pna2_1$, and $Pca2_1$
indicates that the pressure dependence of the $\nu_1$ vibron in these
structures is qualitatively different from that of $Ibam$, and
consistent with the experimental observation of phase V.  $Ibam$ might
become stable at higher pressures, although our DFT calculations do
not support this conjecture.

For completeness, we emphasize that at $0$~K, the metallic atomic
$I4_1/amd$ structure is predicted to become thermodynamically stable
at a pressure of around
$400$~GPa~\cite{prl_dissociation_hydrogen,h_dissociation_morales}.  It
would be interesting to assess the relative stability of $I4_1/amd$
with respect to the mixed structures around room temperature, but this
is beyond the scope of the present work.

Overall, our energetic and spectroscopic results show that $Pca2_1$ is a promising model structure for hydrogen phase V.
It exhibits longer bond lengths compared to those of other similar 
structures, suggesting that phase V is a stepping stone towards the 
metallisation of hydrogen.

\acknowledgements

B.M.\ acknowledges support from the Winton Programme for the Physics of Sustainability, and from Robinson College, Cambridge, and the Cambridge
Philosophical Society for a Henslow Research Fellowship.  E.G.
and R.J.N.\ acknowledge financial support from the Engineering
and Physical Sciences Research Council (EPSRC) of the United Kingdom
(Grants No.\,EP/J003999/1 and No.\,EP/P034616/1,
respectively).  C.J.P.\ is supported by the Royal Society through
a Royal Society Wolfson Research Merit award. The calculations were
performed on the Darwin Supercomputer of the University of Cambridge
High Performance Computing Service facility
(http://www.hpc.cam.ac.uk/), the Archer facility of the UK national
high performance computing service, for which access was obtained via
the UKCP consortium and funded by EPSRC grant No.\ EP/P022596/1, and 
the Oak Ridge Leadership Computing Facility at the
Oak Ridge National Laboratory, which is supported by the Office of Science of the US Department of Energy under Contract No.\,DE-AC05-00OR22725.


\vspace{0.5cm}





 \bibliography{/Users/bartomeumonserrat/Documents/research/papers/references/anharmonic}

\onecolumngrid
\clearpage
\begin{center}
\textbf{\large Supplemental Material for ``Structure and metallicity of phase V of hydrogen''}
\end{center}
\setcounter{equation}{0}
\setcounter{figure}{0}
\setcounter{table}{0}
\setcounter{page}{1}
\makeatletter
\renewcommand{\theequation}{S\arabic{equation}}
\renewcommand{\thefigure}{S\arabic{figure}}
\renewcommand{\bibnumfmt}[1]{[S#1]}
\renewcommand{\citenumfont}[1]{S#1}

\section{Computational details of structure searches}

The structure searches have been performed using the sp-AIRSS method
introduced in this work.  The best candidates known so far for phases
III and IV of hydrogen are monoclinic structures, and therefore our
sp-AIRSS searches have focused on higher-symmetry structures, starting
with orthorhombic crystal systems.

Searches have been performed for simulation cells containing up to
$96$ atoms, at pressures of $100$, $120$, $150$, $200$, $300$, $350$,
$450$, $500$, and $600$~GPa.  The electronic energies were evaluated
using the density functional theory (DFT) plane-wave pseudopotential code {\sc
  castep}~\cite{CASTEP} (version 7) with default ``on the fly''
ultrasoft pseudopotentials~\cite{PhysRevB.41.7892} and the BLYP~\cite{blyp_exchange,blyp_correlation} and PBE~\cite{PhysRevLett.77.3865} density functionals.
Energy cut-offs of $230$~eV, and Monkhorst-Pack $\mathbf{k}$-point grids of spacing $2\pi\times0.07$~\AA$^{-1}$ have been used for the searches. The total number of structures generated by the sp-AIRSS calculations was $45,344$.

\section{Computational details of anharmonic calculations}

The sp-AIRSS calculations found several new structures
with low static lattice enthalpies. These have
been further investigated using the vibrational self-consistent field
method described in Ref.~\cite{PhysRevB.87.144302}. These calculations
consist of two steps: (i) a mapping of the Born-Oppenheimer (BO)
energy surface beyond the local harmonic region, and (ii) the solution
of the resulting vibrational Schr\"{o}dinger equation using a
mean-field ansatz and second order perturbation theory.

The sampling of the BO energy surface follows the principal axes
approximation (PAA)~\cite{jung:10332,PhysRevB.87.144302}:
\begin{equation}
\epsilon_{\mathrm{PAA}}(\mathbf{u})=\epsilon(\mathbf{0}) + \sum_{\mathbf{q},\nu}V_{\mathbf{q}\nu}(u_{\mathbf{q}\nu})+\frac{1}{2}\sum_{\mathbf{q},\nu}\sum_{\mathbf{q}',\nu'}\!{}^{'}V_{\mathbf{q}\nu;\mathbf{q}'\nu'}(u_{\mathbf{q}\nu},u_{\mathbf{q}'\nu'})+\cdots, \label{eq:paa}
\end{equation}
where $u_{\mathbf{q}\nu}$ is the amplitude of a normal mode coordinate
at the vibrational Brillouin zone point $\mathbf{q}$ and branch index
$\nu$, and $\mathbf{u}$ is a vector including the amplitudes of all
normal mode coordinates in the system. The $1$-body terms
$V_{\mathbf{q}\nu}$ are allowed to have a dependence on
$u_{\mathbf{q}\nu}$ beyond that of the harmonic approximation, and the
$2$-body terms provide additional anharmonic corrections arising from
the $2$-dimensional subspaces that they span. The mapping has been
performed with DFT, using the {\sc castep} code~\cite{CASTEP} (version
7) with default ``on the fly'' ultrasoft
pseudopotentials~\cite{PhysRevB.41.7892} and the
BLYP~\cite{blyp_exchange,blyp_correlation} and
PBE~\cite{PhysRevLett.77.3865} functionals. We also exploited the
recently introduced non-diagonal supercells method which allows us to
obtain results that are converged with respect to the simulation cell
size~\cite{non_diagonal}. The computational parameters for the final
calculations reported in the main text consist of an energy cut-off of
$1000$~eV and Monkhorst-Pack $\mathbf{k}$-point grids of spacing
$2\pi\times0.025$~\AA$^{-1}$. These parameters allow us to obtain
differences between frozen-phonon structures that are converged to
better than $10^{-4}$~eV/atom for the energy, forces to better than
$10^{-4}$~eV/\AA, and stresses to better than $10^{-3}$~GPa.  The
anharmonic mapping of the BO energy surface is performed along the
directions determined by the normal modes of vibration, with maximum
amplitudes of $5\sqrt{\langle u^2\rangle}$, and converged results are
obtained by sampling $17$ points along each direction.  The resulting
energy data is fitted using cubic splines, which provide more accurate
fits than polynomial expansions for the strongly anharmonic hydrogen
systems studied here. We note that this differs from the approach used
originally in Ref.~\cite{PhysRevB.87.144302}, in which polynomials
were used to fit the anharmonic BO energy surface.

\begin{figure}
\centering
\subfloat[][One-body term]{\includegraphics[scale=0.9]{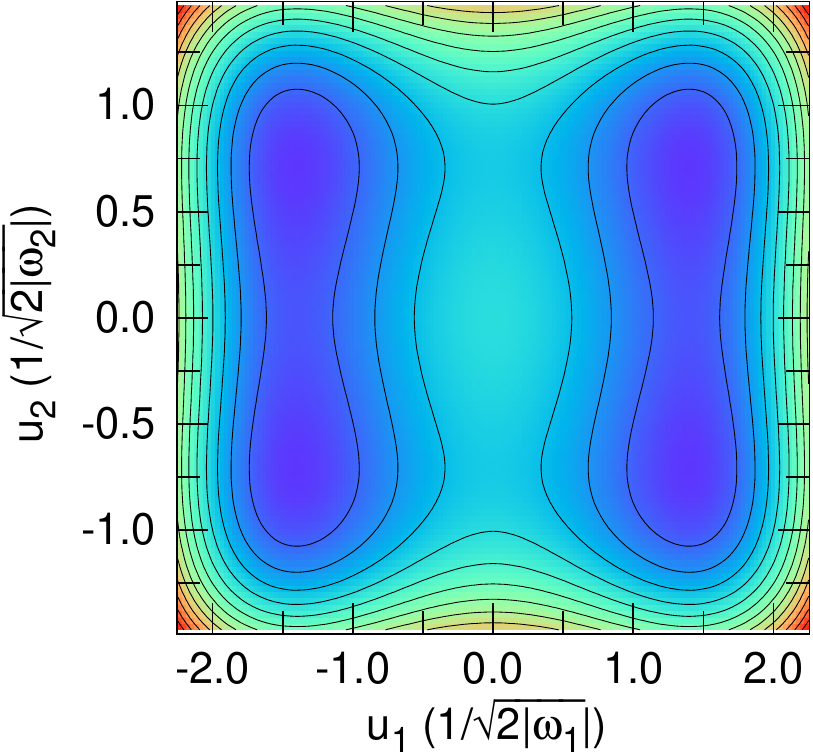}\label{subfig:contour_indep}}
\hspace{0.02cm}
\subfloat[][Two-body term]{\includegraphics[scale=0.9]{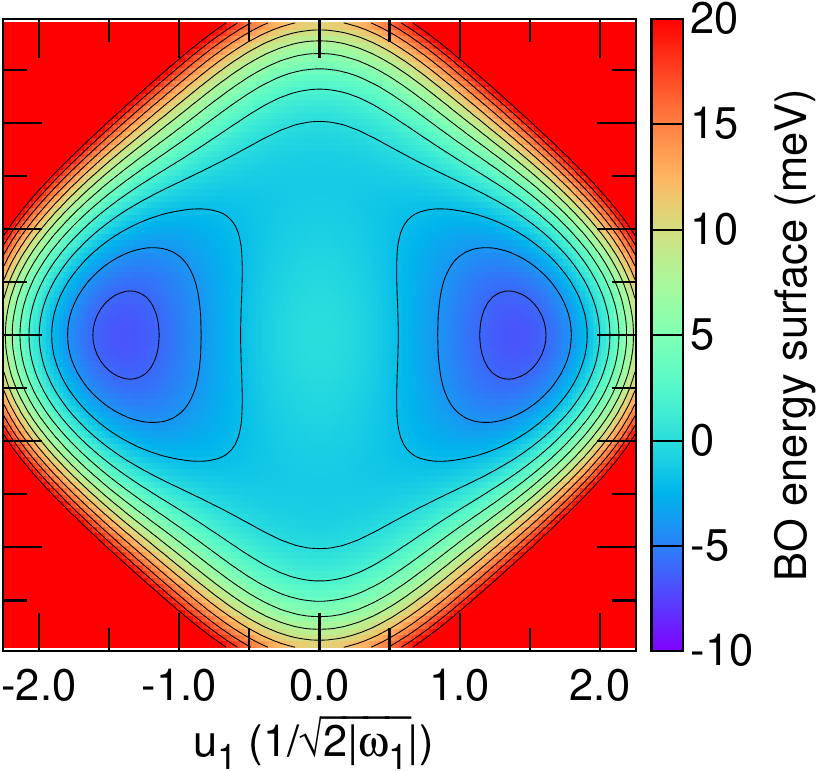}\label{subfig:contour_coupled}}
\caption{$2$-dimensional subspace spanned by two unstable modes within
  (a) the mapping of the BO energy surface along the two
  $1$-dimensional lines $(u_1,0)$ and $(0,u_2)$, and (b) the exact
  mapping of the BO energy surface in the $2$-dimensional plane
  $(u_1,u_2)$. The results are for the $Pca2_1$ structure at
  $P=300$~GPa.}
\label{fig:contour}
\end{figure}

In Fig.~\ref{fig:contour} we show a $2$-dimensional subspace of the BO
energy surface of $Pca2_1$. The subspace is spanned by two normal
modes, $u_1$ and $u_2$, for which the corresponding harmonic frequencies
are imaginary
$\omega_1^2,\omega_2^2<0$. Figure~\ref{subfig:contour_indep} shows the
approximate $2$-dimensional subspace when only $1$-body anharmonic
terms are considered in the expansion of the BO energy surface, while
Fig.~\ref{subfig:contour_coupled} shows the corresponding subspace
with the inclusion of $2$-body terms. The inclusion of $2$-body terms
raises the vibrational energy, therefore contributing to the
stabilisation of the structure.
The inclusion of $2$-body terms affects the energies
of the structures, but does not modify their nature (whether they are
dynamically stable or not.)

The resulting Schr\"{o}dinger equation is solved using a mean-field
ansatz by writing the wave function as a Hartree product, obtaining
the vibrational self-consistent field equations that have been used
previously in molecules and
solids~\cite{bowman:608,PhysRevB.87.144302}:
\begin{eqnarray}
&&\left(-\frac{1}{2}\frac{\partial^2}{\partial u_{\mathbf{q}\nu}^2}+\overline{V}_{\mathbf{q}\nu}(u_{\mathbf{q}\nu})\right)|\phi_{\mathbf{q}\nu}(u_{\mathbf{q}\nu})\rangle=\lambda_{\mathbf{q}\nu}|\phi_{\mathbf{q}\nu}(u_{\mathbf{q}\nu})\rangle, \\
&&\overline{V}_{\mathbf{q}\nu}(u_{\mathbf{q}\nu})\!=\!\left\langle\prod_{\mathbf{q}',\nu'}\!\!{}^{'}\phi_{\mathbf{q}'\nu'}(u_{\mathbf{q}'\nu'})\right|\!\epsilon_{\mathrm{PAA}}(\mathbf{u})\left|\prod_{\mathbf{q}',\nu'}\!\!{}^{'}\phi_{\mathbf{q}'\nu'}(u_{\mathbf{q}'\nu'})\right\rangle.
\end{eqnarray}
The final energy is
\begin{equation}
E=\sum_{\mathbf{q},\nu}\lambda_{\mathbf{q}\nu} + \left\langle\prod_{\mathbf{q},\nu}\phi_{\mathbf{q}\nu}(u_{\mathbf{q}\nu})\right|\epsilon_{\mathrm{PAA}}(\mathbf{u})-\!\sum_{\mathbf{q},\nu}\overline{V}_{\mathbf{q},\nu}(u_{\mathbf{q}\nu})\left|\prod_{\mathbf{q},\nu}\phi_{\mathbf{q}\nu}(u_{\mathbf{q}\nu})\!\right\rangle.
\end{equation}
The second order perturbative correction to the energy in state
$\mathbf{S}$ is given by~\cite{PhysRevB.87.144302}:
\begin{align}
E^{(2)}_{\mathrm{vib},\mathbf{S}}=\sum_{\mathbf{S}'\neq\, \mathbf{S}}\frac{1}{E_{\mathbf{S}}-E_{\mathbf{S}'}}\left|\left\langle\prod_{\mathbf{q},\nu}\phi_{\mathbf{q}\nu}^{S'_{\mathbf{q}\nu}}\right| \epsilon_{\mathrm{PAA}}(\mathbf{u})-\sum_{\mathbf{q},\nu}\overline{V}_{\mathbf{q}\nu}(u_{\mathbf{q}\nu})\left|\prod_{\mathbf{q},\nu}\phi_{\mathbf{q}\nu}^{S_{\mathbf{q}\nu}}\right\rangle\right|^2. 
\end{align}
These energies are then used to calculate the anharmonic free energy $\mathcal{F}$ at finite temperature $T$ according to~\cite{PhysRevB.87.144302}:
\begin{equation}
\mathcal{F}=-\frac{1}{\beta}\ln\mathcal{Z},
\end{equation}
where $\beta=1/k_{\mathrm{B}}T$ is the inverse temperature,
$k_{\mathrm{B}}$ is Boltzmann's constant, and
$\mathcal{Z}=\sum_{\mathbf{S}}e^{-\beta E_{\mathbf{S}}}$ is the
partition function, calculated as a sum over vibrational states
$\mathbf{S}$ of anharmonic energy $E_{\mathbf{S}}$.

We expand the anharmonic wave function associated with each degree of
freedom, $|\phi_{\mathbf{q}\nu}(u_{\mathbf{q}\nu})\rangle$, in terms of a
basis of simple harmonic oscillator eigenstates. Tests show that
converged results are achieved by including $50$ basis functions per
mode. In order to assess the validity of the mean-field formulation,
we consider second-order perturbation theory corrections to the free
energy arising from the subspaces spanned by the soft normal mode
directions, but the resulting energies do not change the stability of
the structures considered.

The reported anharmonic vibrational energies have been calculated
using supercells containing $96$ atoms for $Pc$, $Pna2_1$, $Pca2_1$,
and $Pcaa$. The use of non-diagonal supercells means that the results
are equivalent to those obtained with diagonal supercells containing
$384$ atoms. For $C2/c$ we have used $48$-atom cells, equivalent to
diagonal supercells containing $192$ atoms. For $Cmca$ we have used $36$-atom
cells, equivalent to diagonal supercells containing $324$ atoms, and for
$Ibam$ we have used $24$-atom cells, equivalent
to diagonal supercells containing $216$ atoms. Calculations have been
performed at pressures from $300$~GPa to $450$~GPa, in steps of $50$~GPa.

\section{Quantum Monte Carlo calculations}

We have performed selected diffusion Monte Carlo (DMC) calculations~\cite{RevModPhys.73.33} to confirm that the $Pca2_1$ candidate structure is enthalpically competitive. We have used our previously calculated enthalpies for the $C2/c$, $Pc$, $Cmca$-$12$, and $Cmca$-$4$ structures reported in Ref.~\cite{hydrogen_nature_communications}, and we have performed new calculations for the $Pca2_1$ and $Ibam$ structures.
The numerical details of our DMC calculations are the same as those reported in Ref.~\cite{hydrogen_nature_communications}. In summary, our calculations are based on the geometries optimised at the PBE level of DFT, noting that using BLYP geometries instead only changes the relative enthalpies by about $1$~meV/atom~\cite{hydrogen_nature_communications}. We calculate DMC enthalpies using $96$-atom cells and the PBE-DFT pressures, and use twist averaging to reduce single-particle errors~\cite{twist_average} and the extrapolation scheme of Kwee, Zhang, and Krakauer (KZK) to reduce long-range interaction errors~\cite{qmc_kzk_scheme}. In the KZK scheme, the finite size correction is calculated by comparing standard DFT energies to DFT energies of a modified exchange-correlation functional restricted to a finite sized simulation cell. All DMC calculations have been performed using the {\sc casino} package~\cite{qmc_jphys_review}.

\begin{figure}
\centering
\includegraphics[scale=0.35]{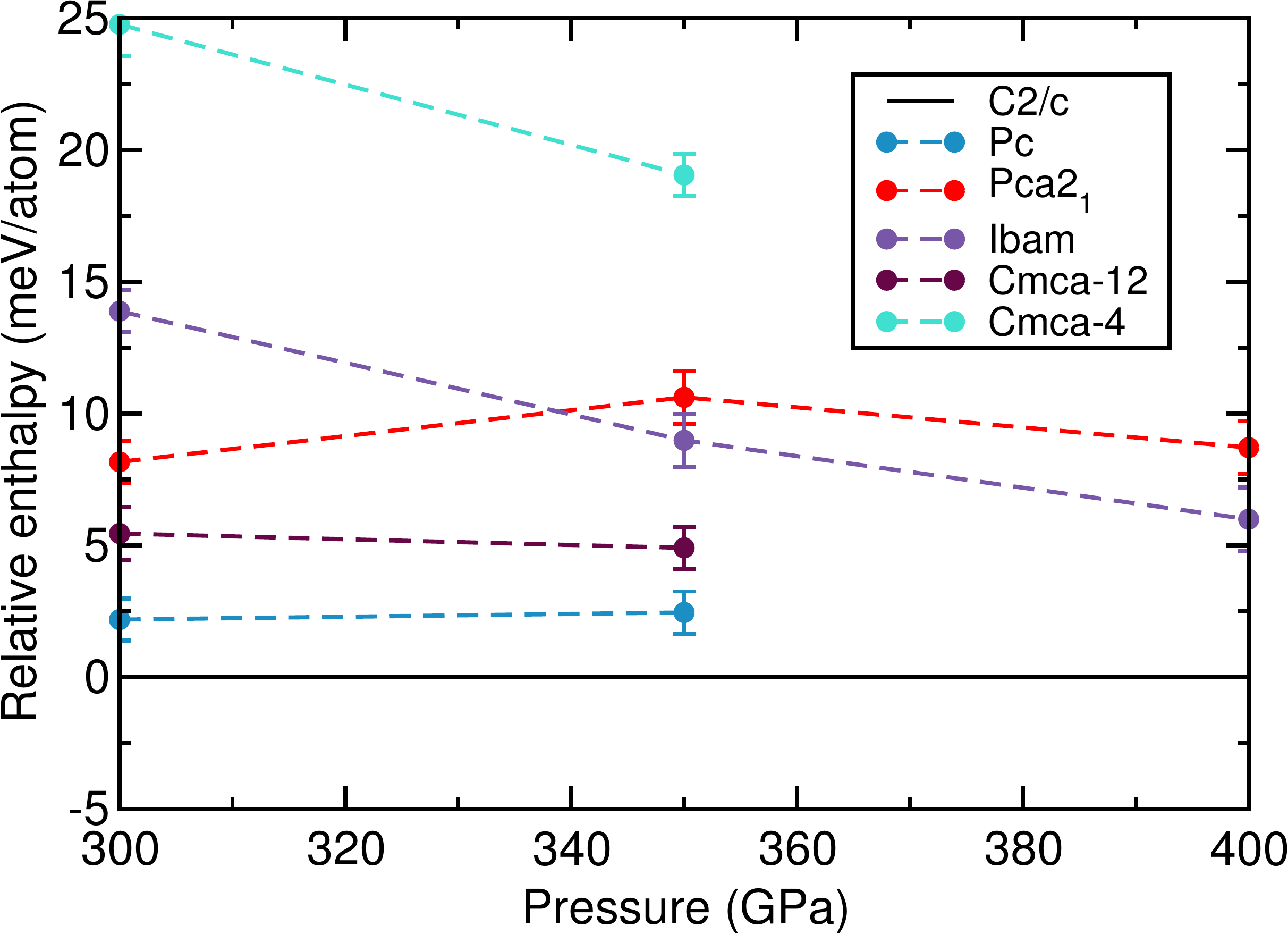}
\caption{Relative enthalpies of selected hydrogen structures calculated using DMC. The dashed lines are only a guide to the eye.}
\label{fig:qmc}
\end{figure}

We reproduce the main text figure again as Fig.~\ref{fig:qmc}, where we report the relative enthalpies of selected candidate structures of high pressure hydrogen. The results for $C2/c$, $Pc$, $Cmca$-$12$, and $Cmca$-$4$ have already been published in Ref.~\cite{hydrogen_nature_communications}. In agreement with these earlier results, we observe that, although $Cmca$-$12$ and $Cmca$-$4$ are enthalpically competitive at the DFT level, this is no longer the case at the DMC level. This trend is further exacerbated by the inclusion of vibrational contributions to the energy, thus justifying the neglect of these two structures in the discussion in the main text. We also observe that $Pc$ is energetically competitive with $C2/c$, and as reported in Ref.~\cite{hydrogen_nature_communications} it is stabilised by thermal motion. This makes the $Pc$ structure a strong candidate for hydrogen phase IV.

The relative enthalpy of the $Ibam$ structure becomes increasingly competitive with increasing pressure, but remains about $5$~meV/atom higher in energy than $C2/c$ at $400$~GPa. Furthermore, the vibrational contribution to the energy is significantly larger in $Ibam$ compared to $C2/c$, thus rendering this structure energetically uncompetitive.

The static lattice enthalpy of the $Pca2_1$ structure at the DMC level is higher than that reported in the main text at the DFT level, and is about $9$~meV/atom higher than that of $C2/c$ at $400$~GPa. However, this difference decreases substantially with the inclusion of the vibrational energy, which at $300$~K is $321$~meV/atom for $C2/c$ and $313$~meV/atom for $Pca2_1$, bringing the relative Gibbs free energy of $Pca2_1$ to only $1$~meV/atom above that of $C2/c$. This suggests, in line with our DFT results, that $Pca2_1$ is energetically competitive at the highest pressures. We note that the $Pc$ structure does not exist above about $375$~GPa, as it falls into the $Cmca$-$4$ structure. This observation, together with the relative Gibbs free energies discussed in this section and the results presented in Ref.~\cite{hydrogen_nature_communications} suggest that $Pc$ is the best candidate for hydrogen phase IV, and $Pca2_1$ is the best candidate for hydrogen phase V. This conclusion is further supported by the comparison of the spectroscopic signatures of the experimental phases and the theoretical structures, as discussed in the main text.

\section{Crystal structures}

\begin{figure} \centering
\subfloat[][$Pc$.]{
\includegraphics[scale=0.070]{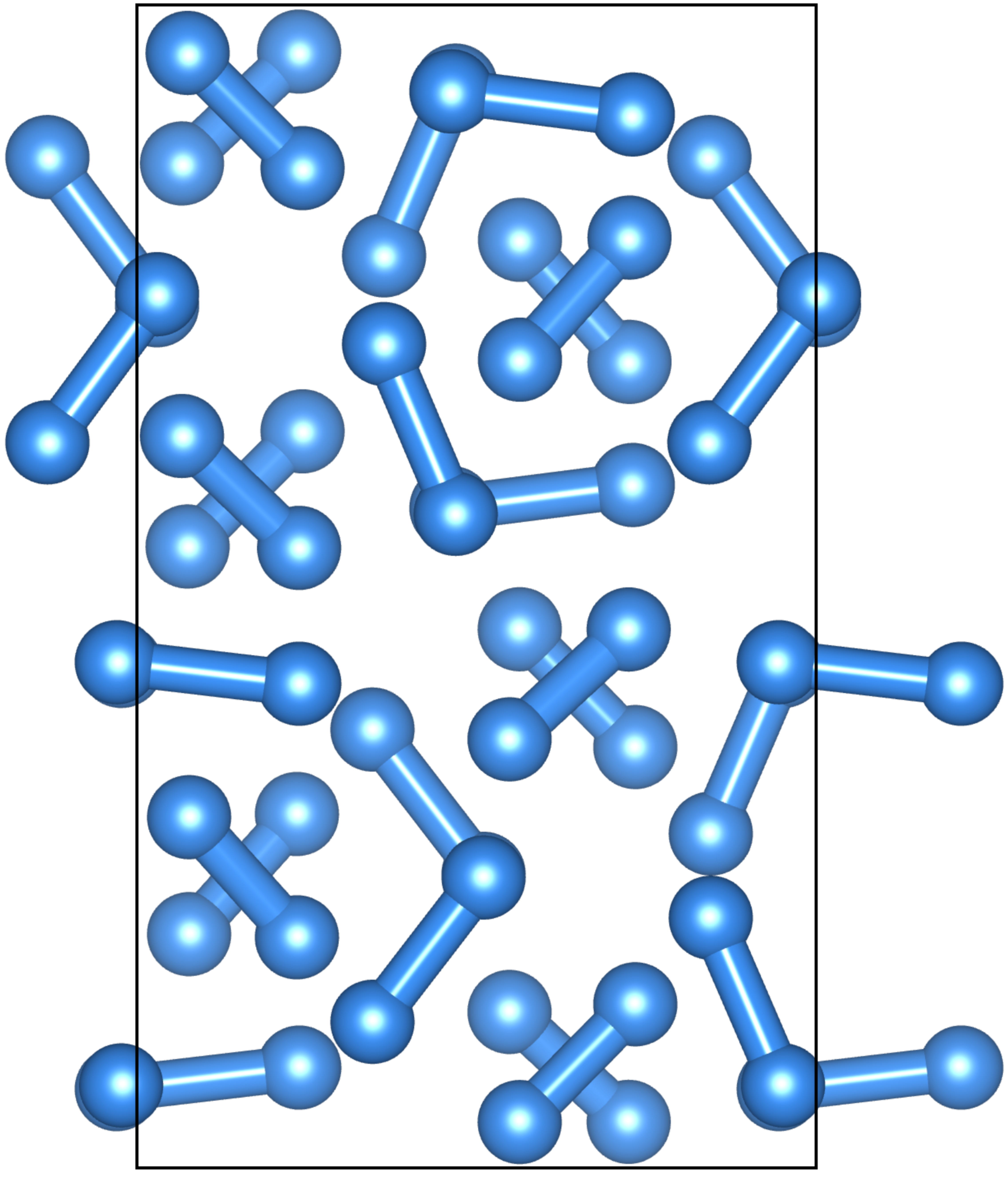}
\includegraphics[scale=0.070]{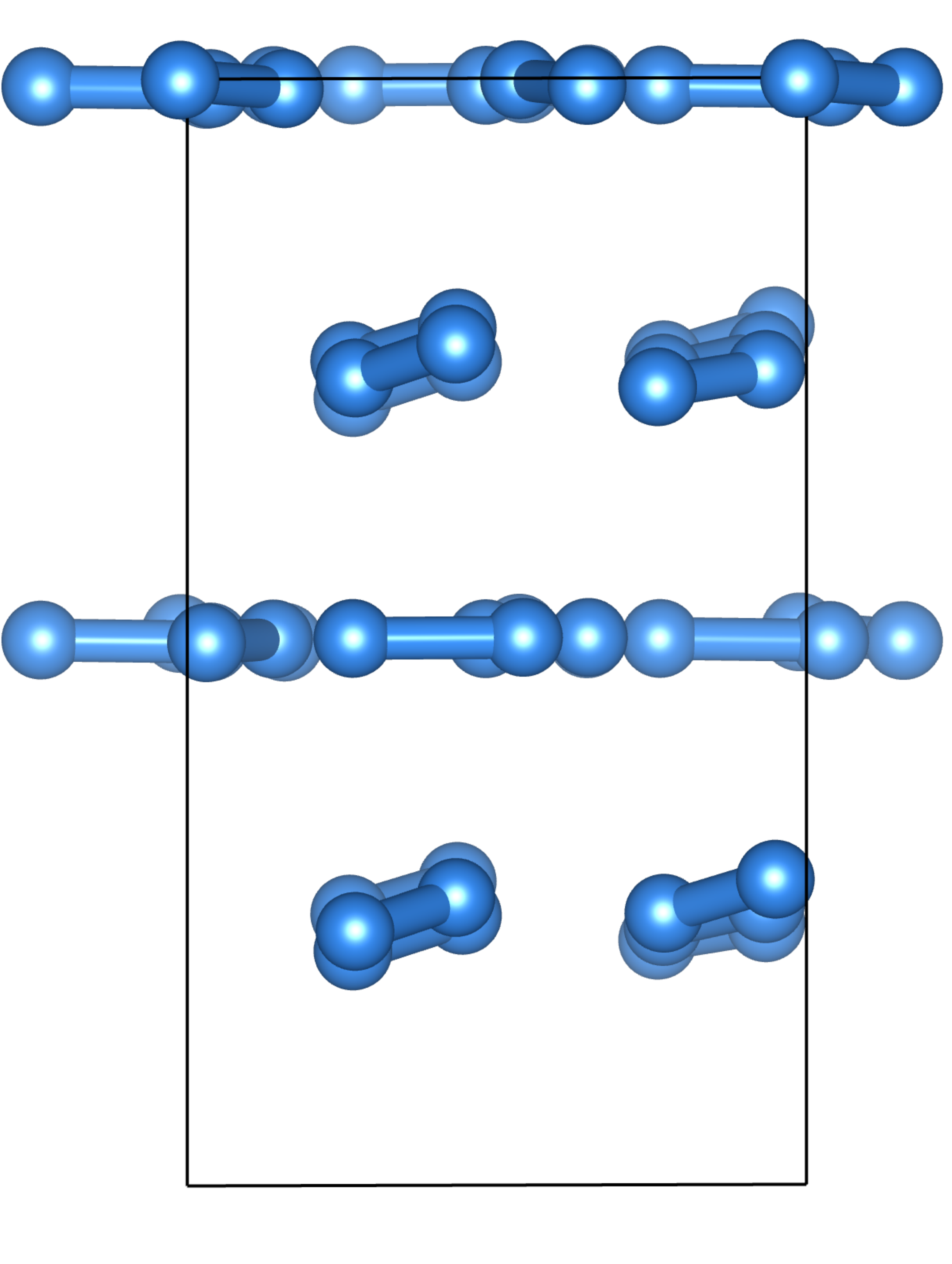}
\label{subfig:Pc}}
\hspace{1.2cm}
\subfloat[][$Pna2_1$.]{
\includegraphics[scale=0.07]{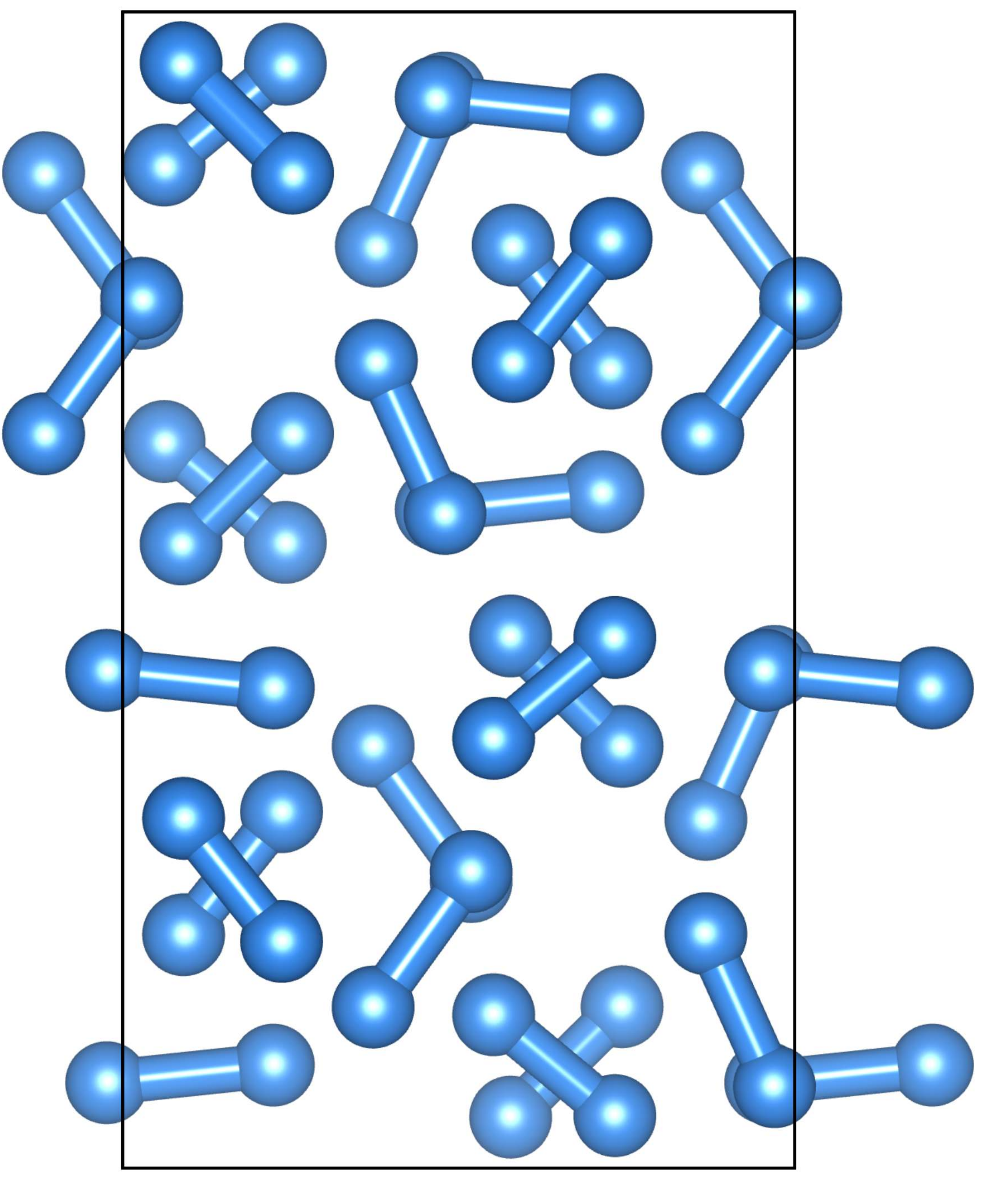}
\includegraphics[scale=0.083125]{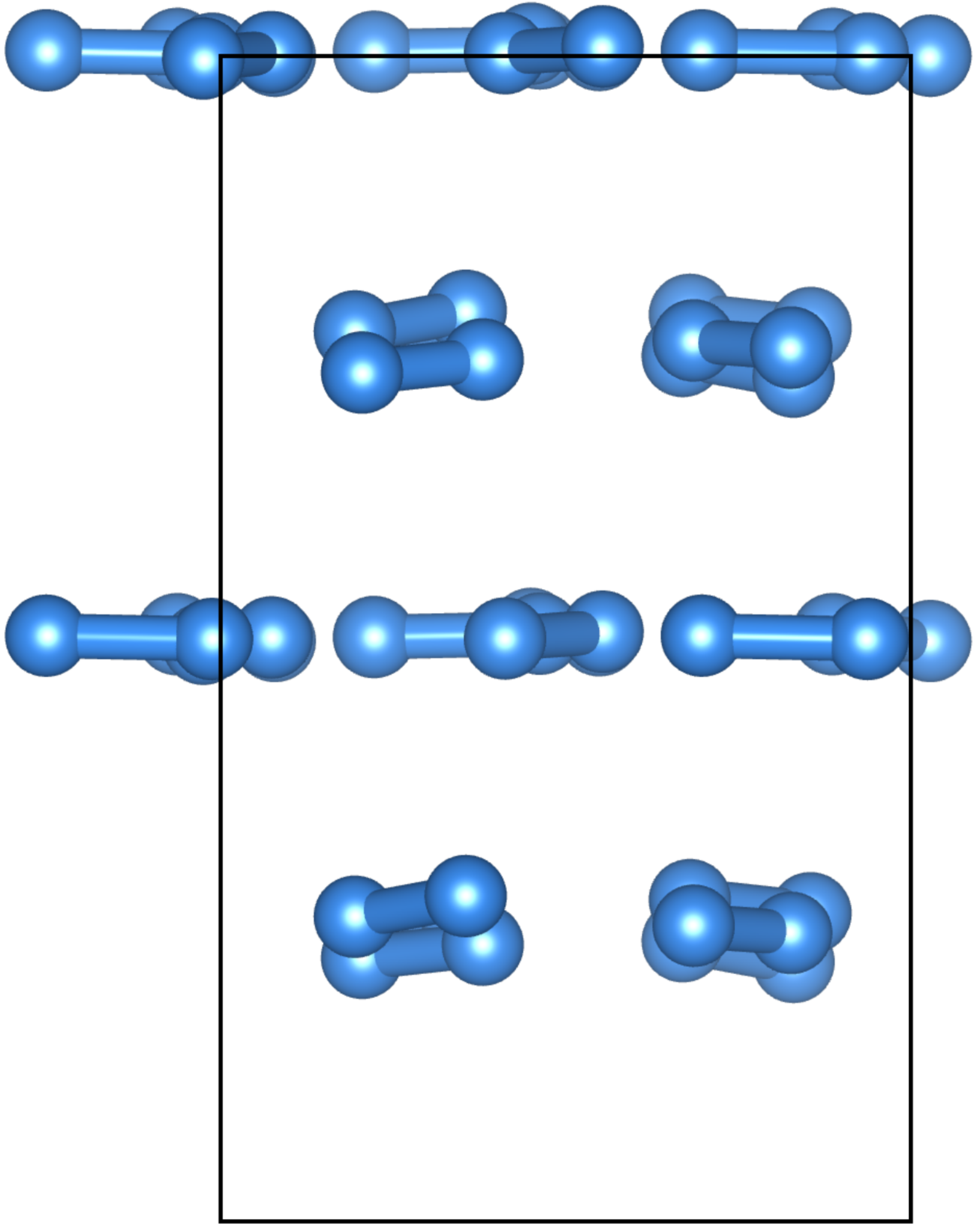}
\label{subfig:Pna21}} \\
\subfloat[][$Pca2_1$.]{
\includegraphics[scale=0.078,angle=90]{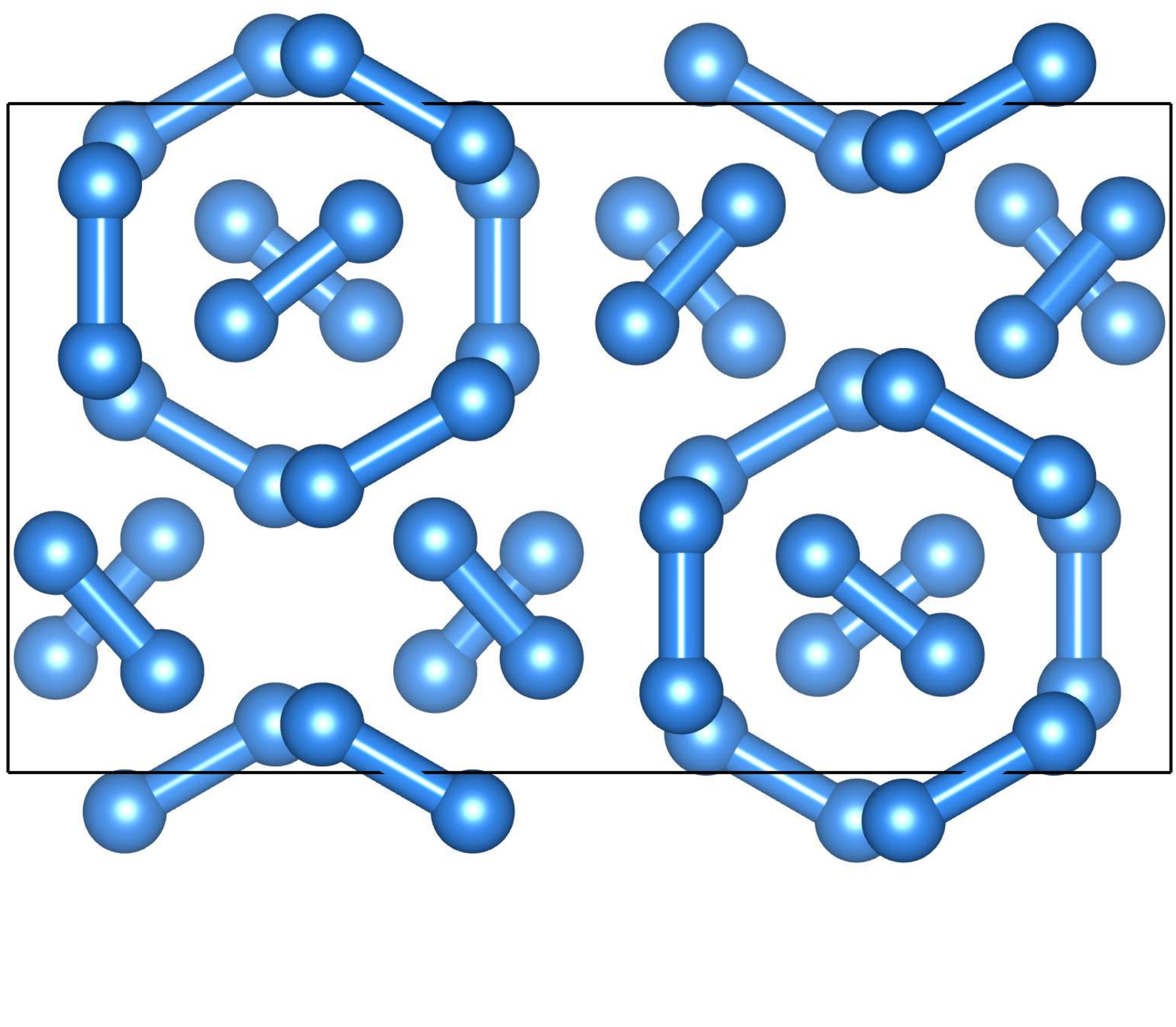}
\includegraphics[scale=0.078]{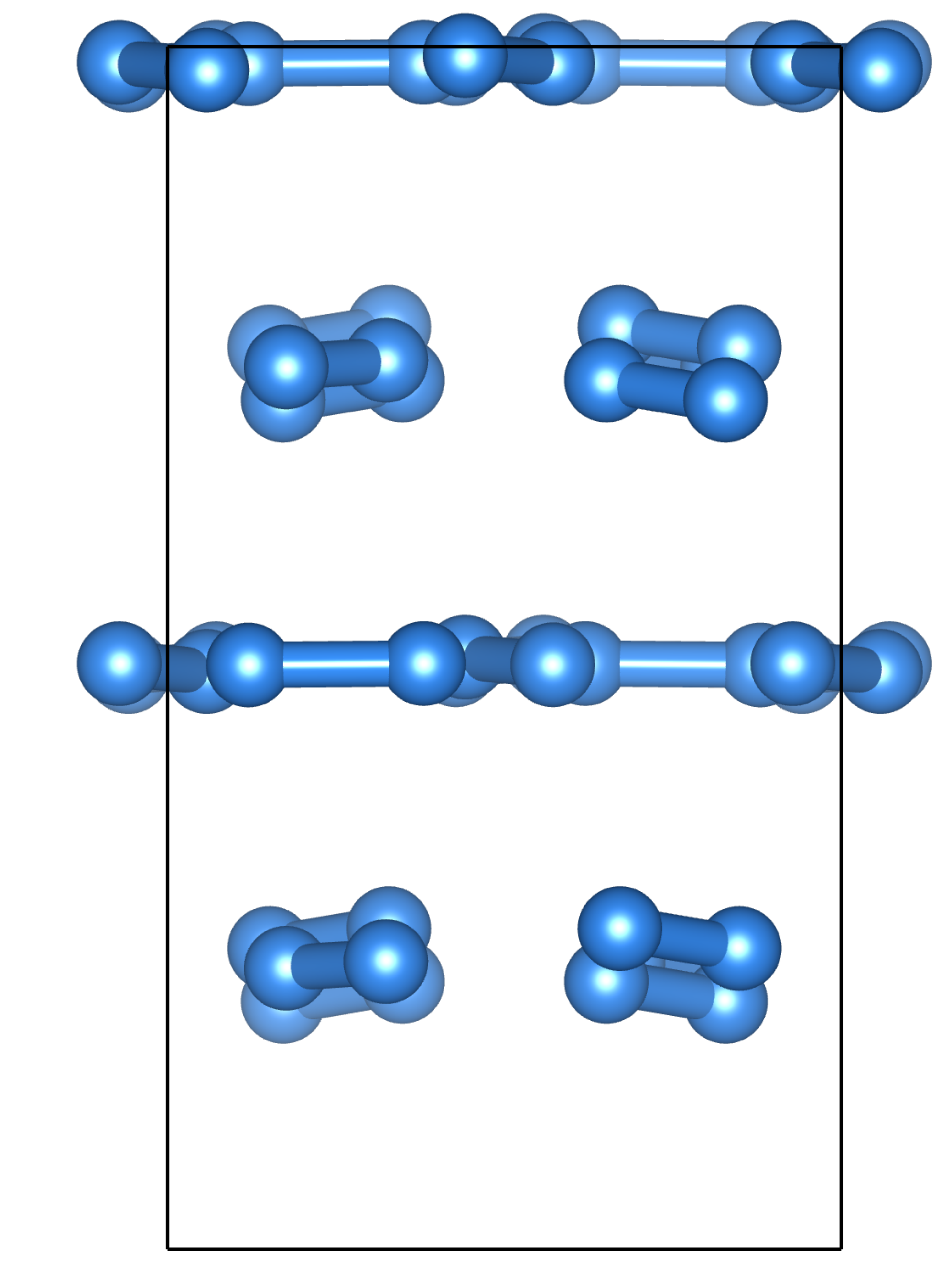}
\label{subfig:Pca21}}
\hspace{1.2cm}
\subfloat[][$Pcaa$.]{
\includegraphics[scale=0.148,angle=90]{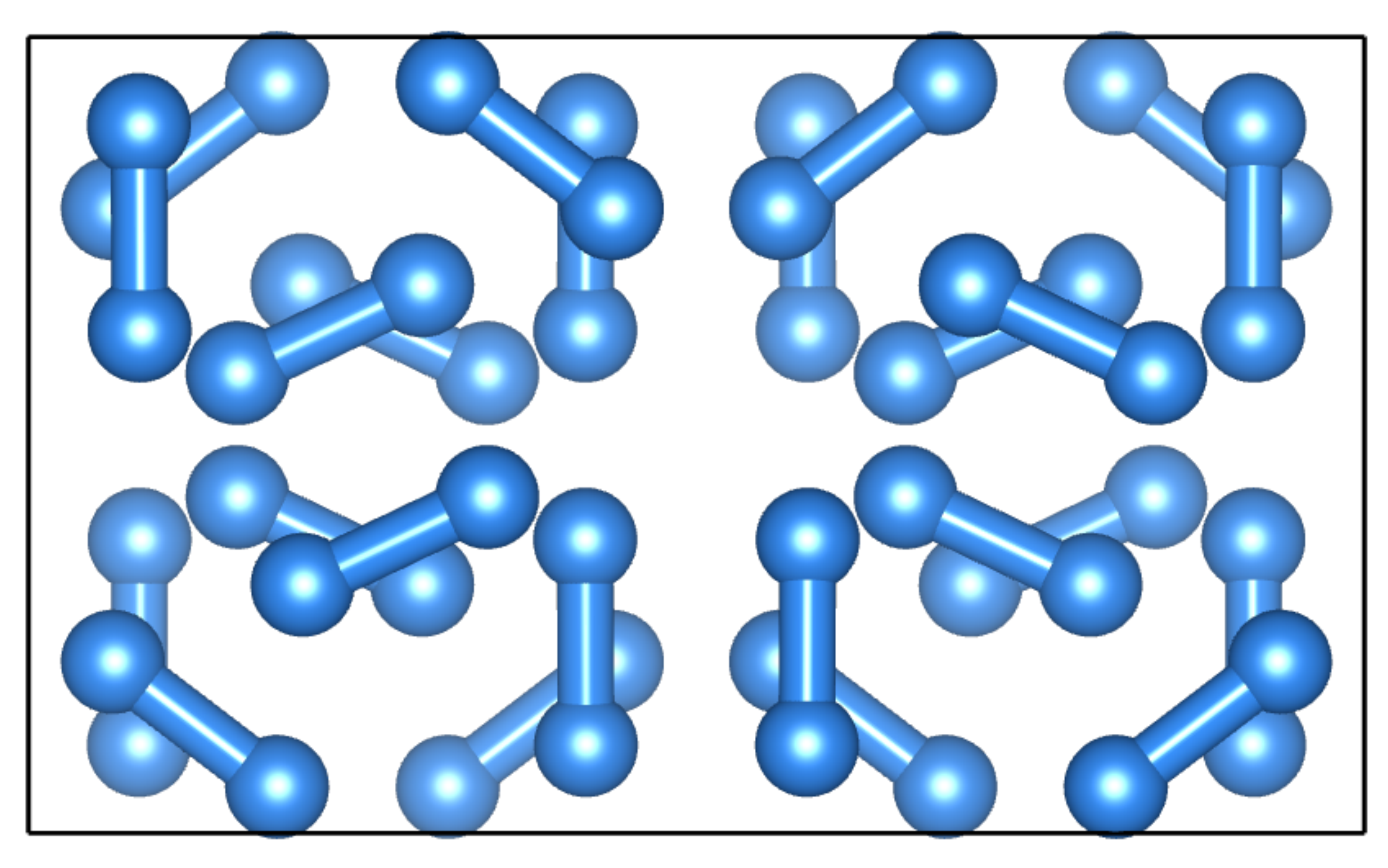}
\hspace{0.2cm}
\includegraphics[scale=0.128]{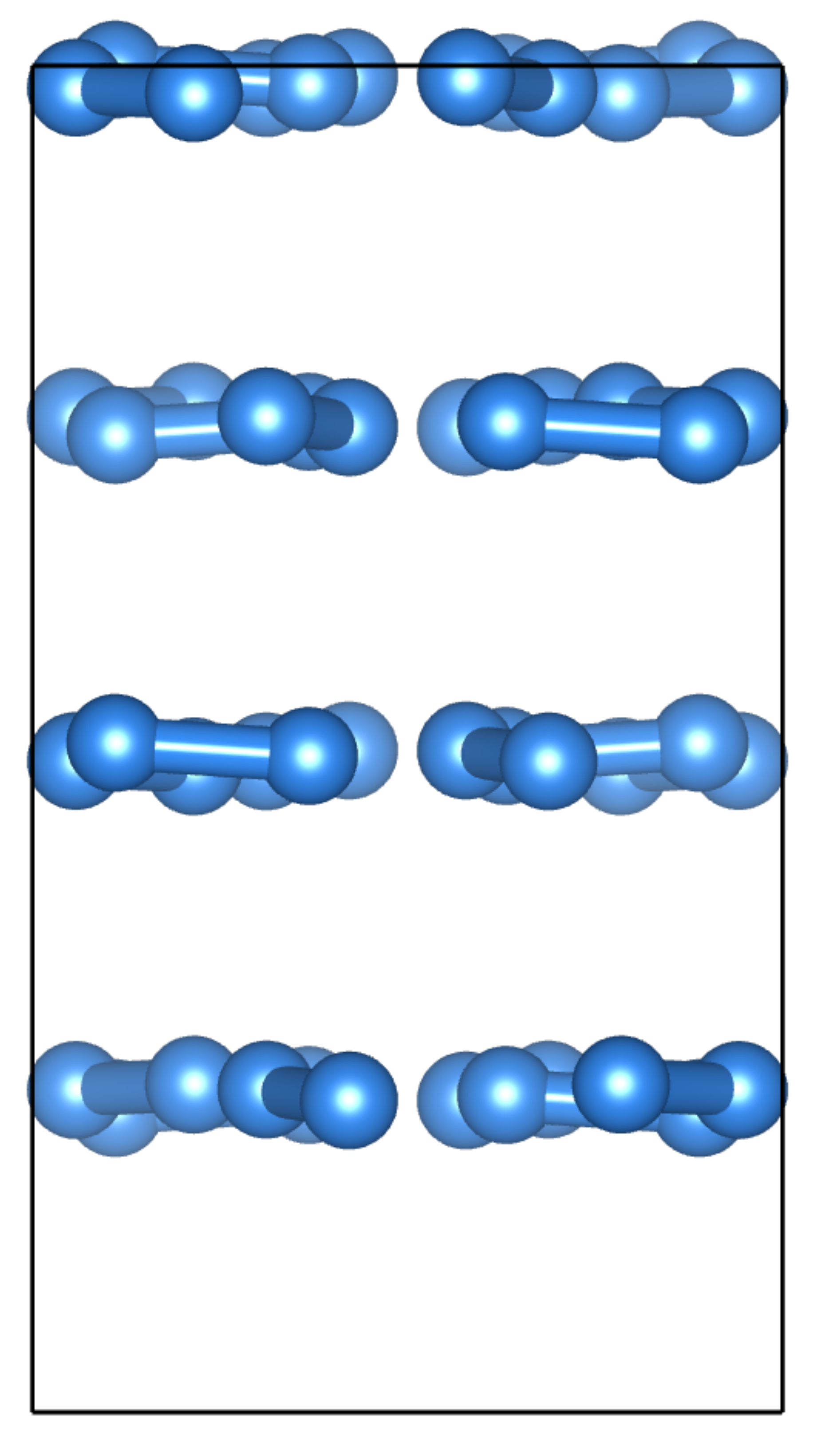}
\label{subfig:Pca21}}
\caption{Top and side views of the $Pc$, $Pna2_1$, $Pca2_1$, and
$Pcaa$ structures.}
\label{fig:structures}
\end{figure}

In Fig.~\ref{fig:structures} we show pictures of the $Pc$, $Pna2_1$, $Pca2_1$
and $Pcaa$ structures. Their layered nature is most clearly seen from the side
view. For $Pc$, $Pna2_1$, and $Pca2_1$, two types layers are distinguishable,
whereas $Pcaa$ exhibits a unique type of layer.

\section{Raman spectra of mixed layered structures}

\begin{figure} \centering
\includegraphics[scale=0.5]{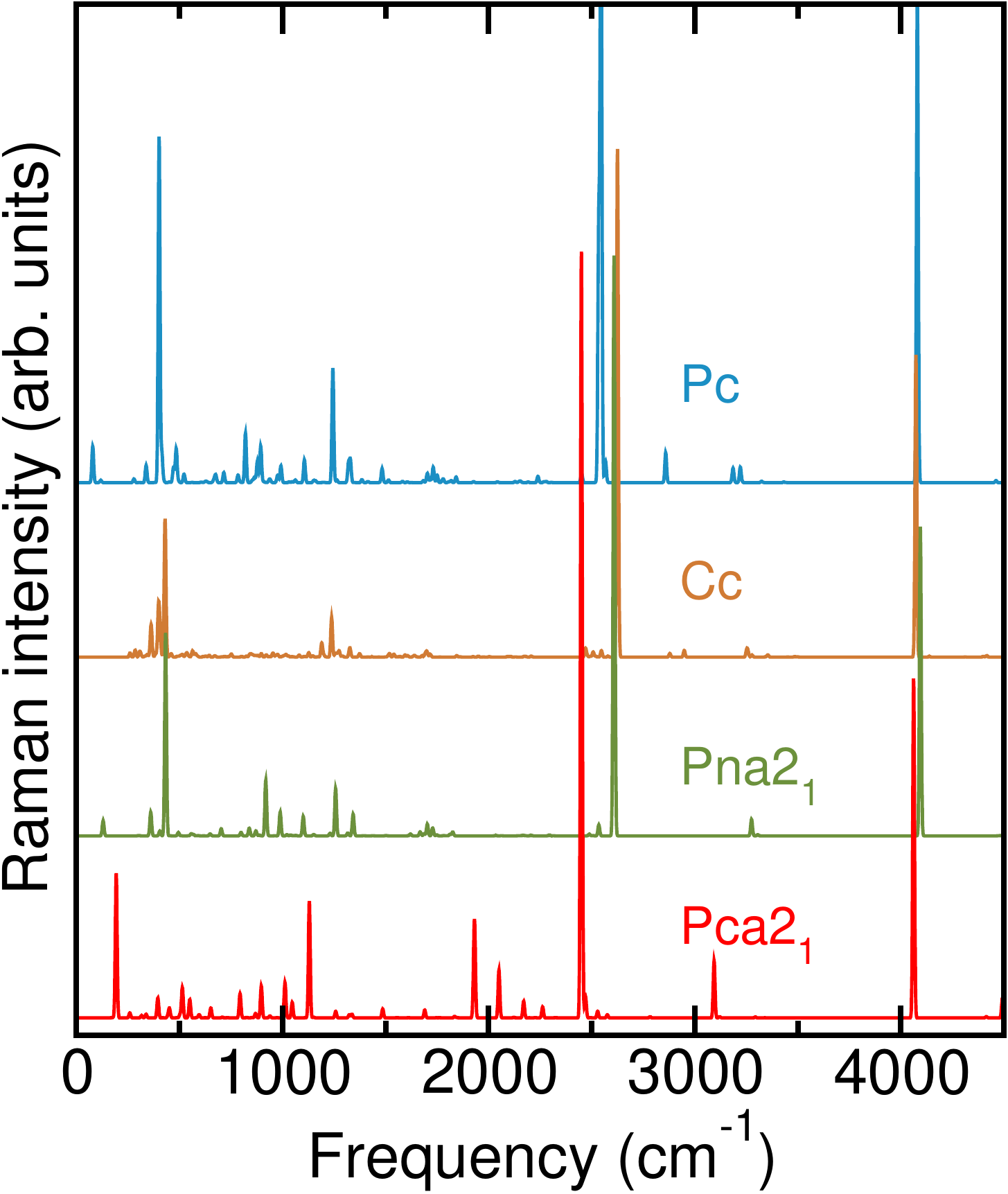}
\caption{Raman spectra of $Pc$, $Cc$, $Pna2_1$, and $Pca2_1$ at
  $374$~GPa.}
\label{fig:raman}
\end{figure}

The Raman spectra reported here and in the main text have been
calculated using the PBE density
functional~\cite{PhysRevLett.77.3865}.  A comparison of the Raman
spectra at $374$~GPa of the four most stable mixed structures of high
pressure hydrogen, $Pc$~\cite{phase_iv_prb},
$Cc$~\cite{ma_cc_hydrogen}, $Pna2_1$, and $Pca2_1$, is shown in
Fig.~\ref{fig:raman}. The Raman spectra of $Pc$, $Cc$, and $Pna2_1$
are almost indistinguishable, and we have used that of the $Pc$ phase
in the main text as representative of the Raman spectra of this set of
structures.

\begin{figure} \centering
\includegraphics[scale=0.48]{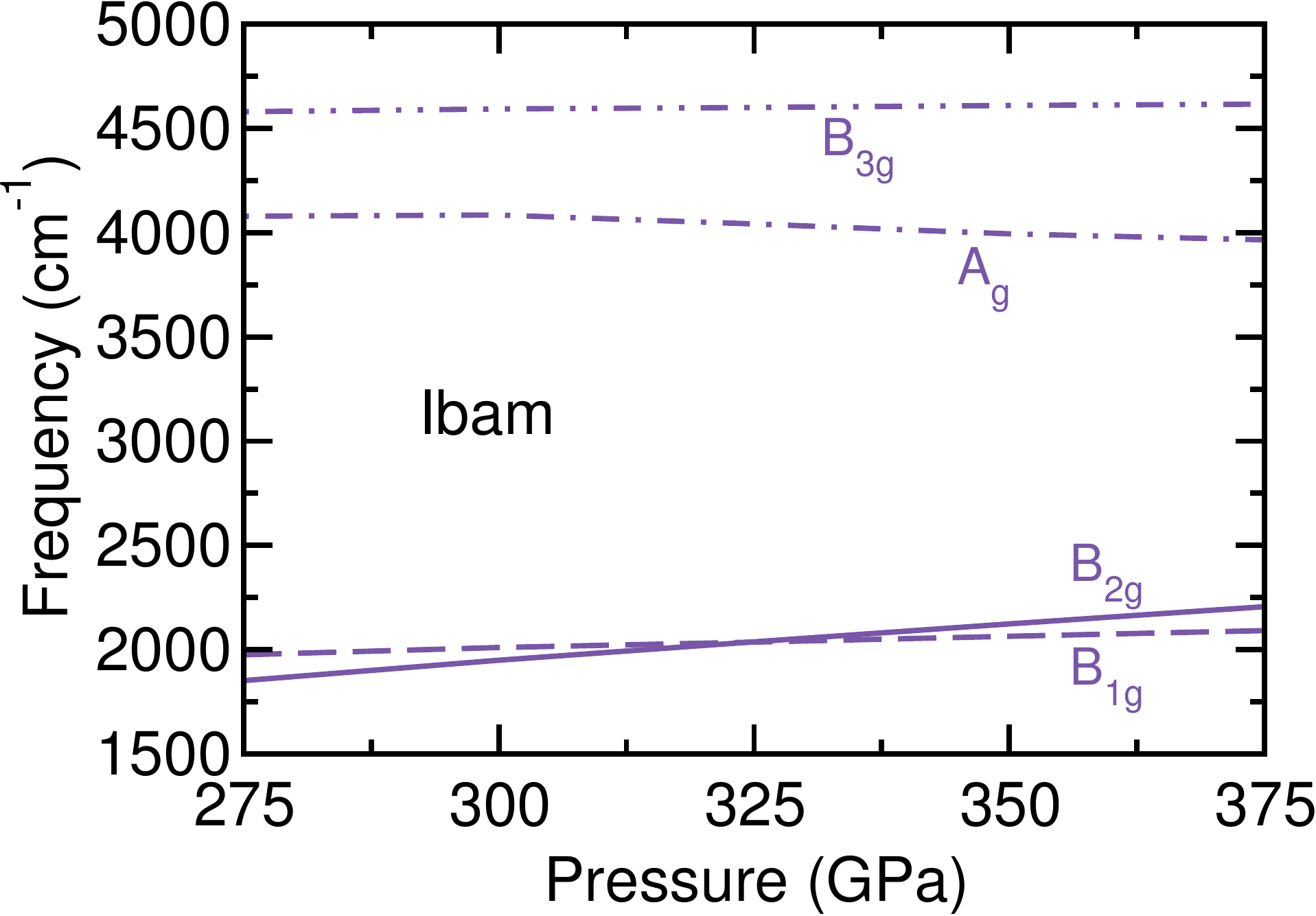}
\caption{Pressure dependence of the frequencies of the Raman active modes of $Ibam$.}
\label{fig:raman_ibam}
\end{figure}

The $Ibam$ structure is metallic within DFT, and as a consequence we
are not able to calculate an accurate Raman intensity
profile. However, we can identify the vibrational modes of $Ibam$ that
are Raman active by considering the symmetries of the system, and we
show the pressue dependence of their frequencies in
Fig.~\ref{fig:raman_ibam}. As described in the main text, the $\nu_1$
vibron frequency increases with pressure as a consequence of the
decrease in bond length in the graphene-like sheets of $Ibam$. This is
inconsistent with the experimental data for phase V.

Our searches have found other new structures which, although not
energetically competitive with $Pca2_1$ at the static lattice level,
have, nonetheless, a relatively low enthalpy. Furthermore, there are
other known candidate structures of hydrogen that are also in this
regime (not necessarily of orthorhombic symmetry or higher). We have
also investigated their Raman spectra, but they are not in agreement
with that observed for phase V. The structures investigated have the
following space groups: monoclinic ($P2/c$, $C2/c$), orthorhombic
($Pcca$, two variants of $Cmca$, $Cccm$, $Fddd$, and two additional
variants of $Ibam$), tetragonal ($P4_32_12$ and $I4_1/amd$), trigonal
($P3_221$), hexagonal ($P6_3/mmc$), and cubic ($Fd\overline{3}m$).

\end{document}